\documentclass[pra,twocolumn,showkeys,showpacs,superscriptaddress,amsmath,amsfonts]{revtex4-1}
\usepackage[utf8x]{inputenc}
\usepackage{microtype}
\usepackage{lmodern}
\usepackage{graphicx}
\usepackage{hyperref}
\usepackage{color}
\usepackage{bm}
\usepackage{nicefrac,xfrac}

\newcommand{\bra}[1]{\langle#1\rvert}
\newcommand{\ket}[1]{\lvert#1\rangle}
\newcommand{\mean}[1]{\langle#1\rangle}
\newcommand{\vect}[1]{\bm{{#1}}}
\begin{document}

\title{Spectral properties of spin-orbital polarons as a fingerprint of orbital order}

\author{Krzysztof Bieniasz}
\email{krzysztof.t.bieniasz@gmail.com}
\affiliation{\mbox{Marian Smoluchowski Institute of Physics, Jagiellonian University,
             Prof. S. {\L}ojasiewicza 11, PL-30348 Krak\'ow, Poland}}
\affiliation{Department of Physics and Astronomy, University of British Columbia,
             Vancouver, British Columbia, Canada V6T 1Z1}
\affiliation{\mbox{Quantum Matter Institute, University of British Columbia,
             Vancouver, British Columbia, Canada V6T 1Z4}}

\author{Mona Berciu}
\affiliation{Department of Physics and Astronomy, University of British Columbia,
             Vancouver, British Columbia, Canada V6T 1Z1}
\affiliation{\mbox{Quantum Matter Institute, University of British Columbia,
             Vancouver, British Columbia, Canada V6T 1Z4}}

\author{Andrzej M. Ole\'s}
\email{a.m.oles@fkf.mpi.de, corresponding author}
\affiliation{Max Planck Institute for Solid State Research,
             Heisenbergstra{\ss}e 1, D-70569 Stuttgart, Germany,}
\affiliation{\mbox{Marian Smoluchowski Institute of Physics, Jagiellonian University,
             Prof. S. {\L}ojasiewicza 11, PL-30348 Krak\'ow, Poland}}

\date{15 May 2019}

\begin{abstract}
Transition metal oxides are a rich group of materials with very
interesting physical properties that arise from the interplay of the
charge, spin, orbital, and lattice degrees of freedom. One interesting
consequence of this, encountered in systems with orbital degeneracy,
is the coexistence of long range magnetic and orbital order, and the
coupling between them. In this paper we develop and study an effective
spin-orbital superexchange model for $e_g^3$ systems and use it to
investigate the spectral properties of a charge (hole) injected into
the system, which is relevant for photoemission spectroscopy. Using an
accurate, semi-analytical, magnon expansion method, we gain insight
into various physical aspects of these systems and demonstrate a number
of subtle effects, such as orbital to magnetic polaron crossover, the
coupling between orbital and magnetic order, as well as the orbital
order driving the system towards one-dimensional quantum spin liquid
behavior. Our calculations also suggest a potentially simple
experimental verification of the character of the orbital order in the
system, something that is not easily accessible through most
experimental techniques.
\end{abstract}

%\pacs{75.25.Dk, 03.65.Ud, 75.10.Lp, 79.60.-i}

\maketitle

\section{Introduction}
\label{sec:intro}

It is a well established fact that the ground state and excitations of
a Hubbard-like model in the regime of strong Coulomb interactions are
faithfully reproduced by an effective model, derived using second order
perturbation theory, which describes almost localized electrons with
suppressed charge fluctuations. The simplest and the most extensively
studied of such models is the $t$-$J$ model~\cite{Cha77}, which
describes an antiferromagnetic (AF) Heisenberg exchange interaction
between localized spins. Doping away from half-filling generates an
electron (or hole) hopping in the subspace without double occupancies,
a formidable many-body problem. Notably, this model predicts that a
charge added to the system will produce a string of misaligned spins,
when the N\'eel AF state is considered, that would trap it in a linear
string potential~\cite{Tru88,Liu92,Man07,Gru18}, while on the other
hand it allows for coherent charge propagation by means of spin
fluctuations~\cite{Kan89,Mar91,Lee06} which remove the spin
excitations produced by the charge. As such, this is a simple
demonstration of a quasiparticle (QP), in which the charge can only
move freely if it couples to the magnetic background of the system.

In systems with active orbital degrees of freedom, such a low-energy
effective model includes superexchange interactions between spins and
orbitals~\cite{Kug82,Tok00}. The development of multiorbital Hubbard
models~\cite{Ole83,Hos16}, most commonly employed in the description 
of transition metal oxides with $d$ orbital degeneracy, led to the
derivation of spin-orbital superexchange models~\cite{Kug82}, which
are $t$-$J$-like model generalizations which accommodate the orbital
degrees of freedom on equal footing with electron spins~\cite{Fei97,Ish97,Fei99,Sna18,Kha00,Kha01,*Kha04,Kha05,Ole05,Cha08,
Hor08,Nor08,*Nor11,*Cha11,Sir08,*Her11,Brz15,Brz19}.
Such models are composed of products of a spin term, characterized by
the common SU(2) symmetry, and the orbital pseudospin part of a lower
symmetry~\cite{Ole05}, reflecting the orbitals' spatial extent and
their interdependence on lattice symmetry. These models allow not only
spin but also orbital long range order in the system, and predict
coherent orbital excitations (orbitons) akin to magnons, to which a
charge can couple in a similar fashion~\cite{vdB00,Ish05,Ishih}.
However, the unusual properties of orbitons and their interaction with
the spin degree of freedom make this problem even more challenging than
the one described above. It is for this reason that these models have
remained a challenge that requires novel theoretical approaches.

Here we are primarily interested in $e_{g}$ systems, which realize a
pseudospin $T=\nicefrac{1}{2}$ interactions and are thus the closest 
analogue of the $t$-$J$ model with $S=\nicefrac{1}{2}$ spins. However, 
due to non-conservation of the orbital quantum number, free propagation 
of charge will be permitted by the kinetic Hamiltonian, and the 
interaction with orbitons will primarily make the resulting QP heavier, 
especially in view of the much smaller role played by orbital 
fluctuations. It was nonetheless suggested that the importance of the 
fluctuations increases with the dimensionality of the $e_g$ problem in 
the case of ferromagnetic (FM) spin order, with one-dimensional (1D) 
alternating orbital (AO) systems being Ising-like~\cite{Dag04}.

On the other hand, for an AF system hole dynamics is dominated by
orbital excitations which leads to quasi-localization when AF and AO
order coexist~\cite{Woh09,Ber09a}. Here we shall address the 
interesting complementary question of what happens in an intermediate
state where AF and AO orders exist simultaneously, but in orthogonal
directions, such that the system can be decomposed into 1D AF chains
and orthogonal two-dimensional (2D) AO planes. Such a situation occurs
in numerous real three-dimensional (3D) systems, in particular in
copper-fluoride perovskite KCuF$_3$~\cite{Oka61}, and in the perovskite
manganite LaMnO$_3$~\cite{Zho06,Kim03}. Both of these systems are of
high interest either from the point of view of basic research, or novel
phenomena triggered by spin-orbital interplay. KCuF$_3$ is a rare
example of a nearly perfect 1D spin liquid~\cite{Lak05,Lak05a}, while
LaMnO$_3$ has almost perfect orbital order and applications stemming
from the colossal magnetoresistance are found in doped
La$_{1-x}$Sr$_x$MnO$_3$~\cite{Jon50,Tok06}.

It is the type of orbital order in spin-orbital systems which is very
intriguing. The orbitals occupied by electrons in LaMnO$_3$ are tuned
by the tetrahedral field which splits the $e_g$ orbitals
\cite{Sna18,Ros19}. It has been realized long ago that the 
photoemission spectra in LaMnO$_3$ strongly depend on the type of 
orbital order in the ground state~\cite{Bal01}, but there is no 
systematic method to measure this order experimentally. Resonance Raman 
spectroscopy~\cite{Kru04} and optical properties~\cite{Kim02,Kov10} 
were proposed to investigate the orbital order but one has to realize 
that the orbitals couple rather strongly to spins~\cite{Sna19} and it 
is thus challenging to investigate the hole coupling to spin-orbital 
excitations in a systematic way. In the regime of intermediate coupling, 
the spectral functions could be obtained using the generalized gradient
approximation with dynamical mean-field theory (GGA+DMFT)~\cite{Kun10}.
Below we use the strong coupling approach and show that the spectral
functions of spin-orbital polarons, obtained from the respective
Green's function, may be used to identify the orbitals occupied in the
ground state.

The remainder of this paper is organized as follows. We introduce the
spin-orbital model with $e_g$ degrees of freedom in Sec.~\ref{sec:model}.
The variational momentum average method used to generate the spectra 
with increasing number of excitations is described in Sec.~\ref{sec:var}. 
In Sec.~\ref{sec:res} we present and discuss the numerical results 
obtained for two representative types of orbital order in the 
intermediate phase with AF/AO order.
The paper is summarized with main conclusions in Sec.~\ref{sec:summa}.
%%AM
%Finally, in the Appendix we present some of the more 
%involved details of the model derivation.
%\textcolor{blue}{
Finally, we present the details of the derivation of the mean field 
phase diagram in Appendix A, and some of the more involved steps of 
the derivation of the fermion-boson polaronic model in Appendix~B.%}

\section{The spin-orbital model}
\label{sec:model}

KCuF$_{3}$ is a tetragonal system (pseudo-cubic to first approximation),
with Cu($d^{9}$) ions placed in octahedral cages of fluorides.
The crystal-field splitting splits the $3d$ orbitals into the
low-lying $t_{2g}$ filled states and the active $e_{g}$ states. Thus,
the copper configuration can be equivalently described as $e_{g}^{3}$
in terms of electron occupation, or $e_{g}^{1}$ in terms of hole
occupation.

The kinetic part of the Hamiltonian includes the electron hopping $t$
between two directional orbitals
\mbox{$\ket{z_{\gamma}}=(3z_{\gamma}^{2}-r_{}^{2})/\sqrt{6}$},
located on nearest neighbor (NN) Cu($3d^{9}$) sites, where
$z_{\gamma}\equiv x/y/z$ is parallel to the main cubic directions
$a/b/c$ of the system~\cite{Fei05}. The complementary orbitals
\mbox{$\ket{\bar{z}_{\gamma}}=(x_{\gamma}^2-y_{\gamma}^2)/\sqrt{2}$}
do not contribute because they are orthogonal to the intermediary 
ligand F($2p^6$) orbitals. The above definition of the hopping is not 
practical, however, due to the orbital basis changing with the hopping 
direction. Transforming all terms into the $\{\ket{z},\ket{\bar{z}}\}$ 
basis we find:
\begin{eqnarray}
  \label{eq:Htso2}
  \mathcal{H}_{t}&=&
  -\frac{t}{4}\sum_{\mean{ij} \perp c}
\left(d_{iz\sigma}^{\dag} \mp\sqrt{3} d_{i\bar{z}\sigma}^{\dag}\right)
\left(d_{jz\sigma}^{}\mp\sqrt{3} d_{j\bar{z}\sigma}^{}\right)\nonumber\\
&-&t\sum_{\mean{ij}\parallel c} d_{iz\sigma}^{\dag}d_{jz\sigma}^{}
+\mathrm{H.c.},
\end{eqnarray}
where the upper/lower sign corresponds to the in-plane directions $a/b$,
respectively. Here, $d_{iz\sigma}^{\dag}$ and $d_{i\bar{z}\sigma}^{\dag}$
create electrons with spin $\sigma$ in the $\ket{z}$ or the
$\ket{\bar{z}}$ orbital, respectively, at site $i$.

The electron interactions are described using a multiorbital
Hubbard-like model, including on-site Coulomb repulsion $U$ and Hund's
exchange interaction $J_H$ which drives the site towards maximal spin.
We are interested in the strongly correlated limit $U \gg t$, which,
when considering virtual excitations,
$e_{g}^{3}e_{g}^{3} \rightleftharpoons e_{g}^{2}e_{g}^{4}$,
leads to an effective superexchange model~\cite{Kug82}. Due to the
proximity of degeneracy of the $e_{g}$ orbitals, one needs to consider
the multiplet structure of the $e_{g}^{2}$ ion. The spectrum of these
excitations has four eigenenergies $U-3J_{H}$, $U-J_{H}$ (double), and
$U+J_H$~\cite{Ole00}. Taking all this into consideration leads to the
following superexchange Hamiltonian:
\begin{subequations}
  \label{eq:Hexch}
  \begin{align}
\mathcal{H}_{1}^{\gamma} &= -2Jr_{1} \sum_{\mean{ij}\parallel\gamma}
\left(\vect{S}_i\cdot \vect{S}_j+\frac{3}{4}\right)
      \left(\frac{1}{4}-\tau_{i}^{\gamma}\tau_{j}^{\gamma}\right),\\
\mathcal{H}_{2}^{\gamma} &= 2Jr_{2} \sum_{\mean{ij}\parallel\gamma}
\left(\vect{S}_i\cdot \vect{S}_j-\frac{1}{4}\right)
      \left(\frac{1}{4}-\tau_{i}^{\gamma}\tau_{j}^{\gamma}\right),\\
\mathcal{H}_{3}^{\gamma} &= 2Jr_{3} \sum_{\mean{ij}\parallel\gamma}
\left(\vect{S}_i\cdot \vect{S}_j-\frac{1}{4}\right)
\left(\frac{1}{2}-\tau_{i}^{\gamma}\right)
\left(\frac{1}{2}-\tau_{j}^{\gamma}\right),\\
\mathcal{H}_{4}^{\gamma} &= 2Jr_{4} \sum_{\mean{ij}\parallel\gamma}
\left(\vect{S}_i\cdot \vect{S}_j-\frac{1}{4}\right)
\left(\frac{1}{2}-\tau_{i}^{\gamma}\right)
\left(\frac{1}{2}-\tau_{j}^{\gamma}\right),
  \end{align}
\end{subequations}
where the $\{r_i\}$ coefficients serve to impose the multiplet
structure at finite Hund's exchange $J_H>0$,
\begin{equation}
  \label{eq:ri}
  r_{1}=\frac{1}{1-3\eta}, \quad
  r_{2}=r_{3}=\frac{1}{1-\eta}, \quad
  r_{4}=\frac{1}{1+\eta},
\end{equation}
with
\begin{equation}
\eta=J_H/U\,,
\label{eta}
\end{equation}
while $\tau_i^{\gamma}$ are bond-direction-dependent orbital operators
for the principal cubic axes, which can be expressed using the 
pseudospin operators in the following way:
\begin{equation}
  \label{eq:tau}
\tau_{i}^{a/b}=-\frac{1}{2}\left(T_i^z\mp\sqrt{3}T_i^x\right), \qquad
  \tau_{i}^{c}=T_{i}^{z},
\end{equation}
under the standard convention,
\begin{equation}
  \label{eq:pseudospin}
  \ket{\bar{z}}\equiv\ket{\uparrow}, \qquad
  \ket{z}\equiv\ket{\downarrow}.
\end{equation}

It can be shown that assuming a FM spin state in the $ab$ planes and 
under a purely octahedral crystal field, the orbital order preferred 
by the superexchange Hamiltonian is AO, with the
\begin{equation}
\label{pm}
\ket{\pm}=(\ket{\bar{z}}\pm\ket{z})/\sqrt{2}
\end{equation}
states occupied. However, this need not be the case for other magnetic
orders. In the general case, the occupied orbitals are given by rotation 
of the basis, which is most conveniently parametrized with an angle 
$\pm(\pi/2+\phi)$, where the sign depends on the orbital sublattice, 
with $\phi=0$ corresponding to the $\{|+\rangle,|-\rangle\}$ reference 
basis (\ref{pm}).

For further convenience, we also introduce an orbital crystal field
into the Hamiltonian, which serves to remove the orbital degeneracy
of the system~\cite{Sna18}, and to make the model more realistic
\cite{Brz12}:
\begin{equation}
  \label{eq:Hz}
  \mathcal{H}_{z} = -E_{z}\sum_{i} T_{i}^{z}.
\end{equation}
This term simulates an axial pressure along the $c$ axis, and for large
values of $\lvert E_{z}\rvert$ it supports ferro-orbital (FO) order,
with occupied states either $\ket{\bar{z}}$ (for $E_z>0$) or $\ket{z}$
(for $E_z<0$). Tuning the orbital field thus allows one to drive the
system from AO all the way to FO order in a continuous manner, 
%\textcolor{blue}{
although we will not be interested in this extreme limit%}. Note that some 
of the superexchange terms are similar to the crystal field in that they
are linear in the $\tau^{\gamma}$ operators, and thus when these are
active (\emph{i.e.}, when the magnetic order is not assumed to be FM)
there is an internal orbital field already present in the superexchange
Hamiltonian. Thus, the external field will work either to counter or to
enhance these terms, in turn affecting the magnetic order. In this way
the system incorporates spin-orbit coupling through indirect means,
allowing for the magnetic and orbital orders to affect each other and,
furthermore, to be controlled through external parameters, such as an
axial pressure.

%\textcolor{blue}{
In order to derive an effective polaronic Hamiltonian for a single 
charge doped into the system, we need to perform a series of rather 
involved steps: 
(i)~determine the classical ground state by calculating the mean field 
energy and minimizing it with respect to the crystal field~$E_{z}$,
for more details see Appendix A; 
(ii)~transform the kinetic part of the Hamiltonian~\eqref{eq:Htso2} to 
the orbital basis corresponding to the classical ground state; 
(iii) introduce magnons and orbitons (to represent magnetic and orbital 
excitations above the classical ground state) as slave bosons by means 
of a Holstein-Primakoff transformation. As these operations are rather 
tedious and unlikely to be of much interest to the general audience, 
we relegate this derivation 
of the polaronic Hamiltonian to Appendix B. It is only 
important to notice that from this point onward we will be mostly 
relying on the outlined formalism, and thus we will be referring to 
magnetic and orbital excitations as magnons (denoted with the operators 
$b_{i}^{\dag}$) and orbitons (denoted as $a_{i}^{\dag}$), 
respectively, and treating them as well-defined, spinless bosons, while 
the charge degree of freedom will be represented by the spinless fermion $f_i^{\dag}$.%}

%\textcolor{blue}{
The final Hamiltonian consists of the exchange term 
%%AM
%$\mathcal{H}_{J}=\mathcal{H}_{\mathrm{I}}+\mathcal{H}_{\mathrm{II}}$, 
%given by Eqs~\eqref{eq:Hphi}, and the kinetic terms 
%$\mathcal{H}_t=\mathcal{T}+\mathcal{V}^{\perp}_t+\mathcal{V}^{\parallel}_t$, 
%defined by Eqs.~\eqref{eq:Htso4}. 
$\mathcal{H}_{J}$, and the kinetic term $\mathcal{H}_t$. 
%defined in the Appendix. 
%
It is important for the understanding 
of the paper what physical processes are realized by each of those 
terms. The exchange term, 
%%AM
%$\mathcal{H}_{J}$ 
$\mathcal{H}_J\equiv\mathcal{H}_{\mathrm{I}}+\mathcal{H}_{\mathrm{II}}$, 
is of course responsible 
for the spin-orbital order in the presence of the crystal field; 
here we have conveniently divided it into the terms quadratic in 
(pseudo)spin operators, included in  $\mathcal{H}_{\mathrm{I}}$, and the linear 
(crystal field like) terms included in $\mathcal{H}_{\mathrm{II}}$,
%%AM
see the Appendix B. 
After the 
Holstein-Primakoff transformation these terms are purely bosonic 
operators, and include the Ising terms which only serve to 
``count'' the bosonic energy, and the fluctuation terms which create 
and destroy the various bosons without involving the doped charge,
%%AM
similar to the spin polaron in the $t$-$J$ model \cite{Mar91}.%}

%\textcolor{blue}{
The kinetic Hamiltonian, 
$\mathcal{H}_t\equiv
\mathcal{T}+\mathcal{V}^{\perp}_t+\mathcal{V}^{\parallel}_t$, 
on the other hand, contains all of the charge dynamics,
%%AM
as shown in the Appendix B. 
The free hopping term $\mathcal{T}$ is restricted to the FM $ab$ 
planes due to spin conservation---any hopping out of plane necessarily 
produces magnons. The  $\mathcal{V}_t$ term includes all 
the processes responsible for the electron-boson coupling and 
constitute the actual interaction in our model. Because of  the in-plane FM order, the perpendicular term, $\mathcal{V}_{t}^{\perp}$, can only 
produce orbitons, while its influence on magnons is limited to a 
fermion-magnon swap term. 
%%AM
Finally,
the out of plane term $\mathcal{V}_{t}^{\parallel}$ 
describes hole dynamics by the coupling
to both magnons and orbitons at the same time. Altogether, 
these terms represent all the fermion-boson coupling processes possible 
in this system and include terms as complicated as five particle 
interactions. Our variational technique, which we will briefly describe 
in the next section, allows us to include all of those terms, something 
that would not be possible to do in  more standard polaronic 
methods relying on the linear spin wave (LSW) approximation.%}

%\textcolor{red}{
It needs to be emphasized, however, that the present model employs a
number of idealizations (\emph{e.g.}, we neglect the intermediary oxygen
orbitals and proper Jahn-Teller interactions, and ignore any resulting
structural transitions that might occur in the system)
and is not intended to produce a realistic low
energy excitation spectrum, but rather to study the effects of spin and
orbital excitations on the charge dynamics in systems with the
$A$-AF/$C$-AO ground state, as encountered in KCuF$_3$ and LaMnO$_3$.
The results presented here are therefore not meant to directly address
the experimental results, although some of the observed qualitative
effects could be relevant to interpret or guide the experiment.%}

\section{The Momentum Average method}
\label{sec:var}

We use the well-established momentum average (MA) variational method
\cite{Ber06,Mar10,Ber11,Ebr15} to determine the one-electron Green's
function,
$G(\vect{k},\omega)=\bra{\vect{k}}\mathcal{G}(\omega)\ket{\vect{k}}$,
where \mbox{$\mathcal{G}(\omega)=[\omega+i\eta-\mathcal{H}]^{-1}$} is 
the resolvent operator and 
\mbox{$\ket{\vect{k}}=f_{\vect{k}}^{\dag}\ket{0}$} is the Bloch state 
for an electron injected into the undoped, semiclassical ground 
state~$\ket{0}$. The Hamiltonian $\mathcal{H}$ is divided into
$\mathcal{H}_{0}=\mathcal{T}+\mathcal{H}_J^z$, where $\mathcal{H}_J^z$
is the Ising part of the exchange terms in Eq.~\eqref{eq:Hphi}
(usually, the quantum fluctuations are of little importance and can be
ignored, see also Ref.~\cite{Bie16}), and the interaction,
$\mathcal{V}=\mathcal{V}^{\perp}_{t}+\mathcal{V}^{\parallel}_{t}$,
which might also be extended to include the spin fluctuation terms of
the exchange Hamiltonian.

The variational MA method uses Dyson's identity,
\begin{equation}
 \mathcal{G}(\omega)=\mathcal{G}_{0}(\omega)
+\mathcal{G}(\omega)\mathcal{V}\mathcal{G}_{0}(\omega),
\end{equation}
to generate the equations of motion (EOMs) for the Green's functions,
within a chosen variational space. Specifically, evaluation of
$\mathcal{V}\ket{\vect{k}}$ in real space links to generalized
propagators that involve various bosons beside the fermion; the
variational expansion controls which such configurations are included
in the calculation. The EOMs for these generalized Green's functions
are then obtained using the same procedure and the process is continued
until all the variational configurations are exhausted, at which point
this hierarchy of coupled EOMs automatically truncates. The validity
and accuracy of the approximation is determined by how appropriate is
the choice of the variational space; this is usually based on some
physically-motivated criterion restricting the spatial spread of the
bosonic cloud, as exemplified below. The accuracy of the results can
be systematically improved by increasing the variational space until
convergence is achieved.

\begin{figure}[t!]
  \includegraphics[width=\columnwidth]{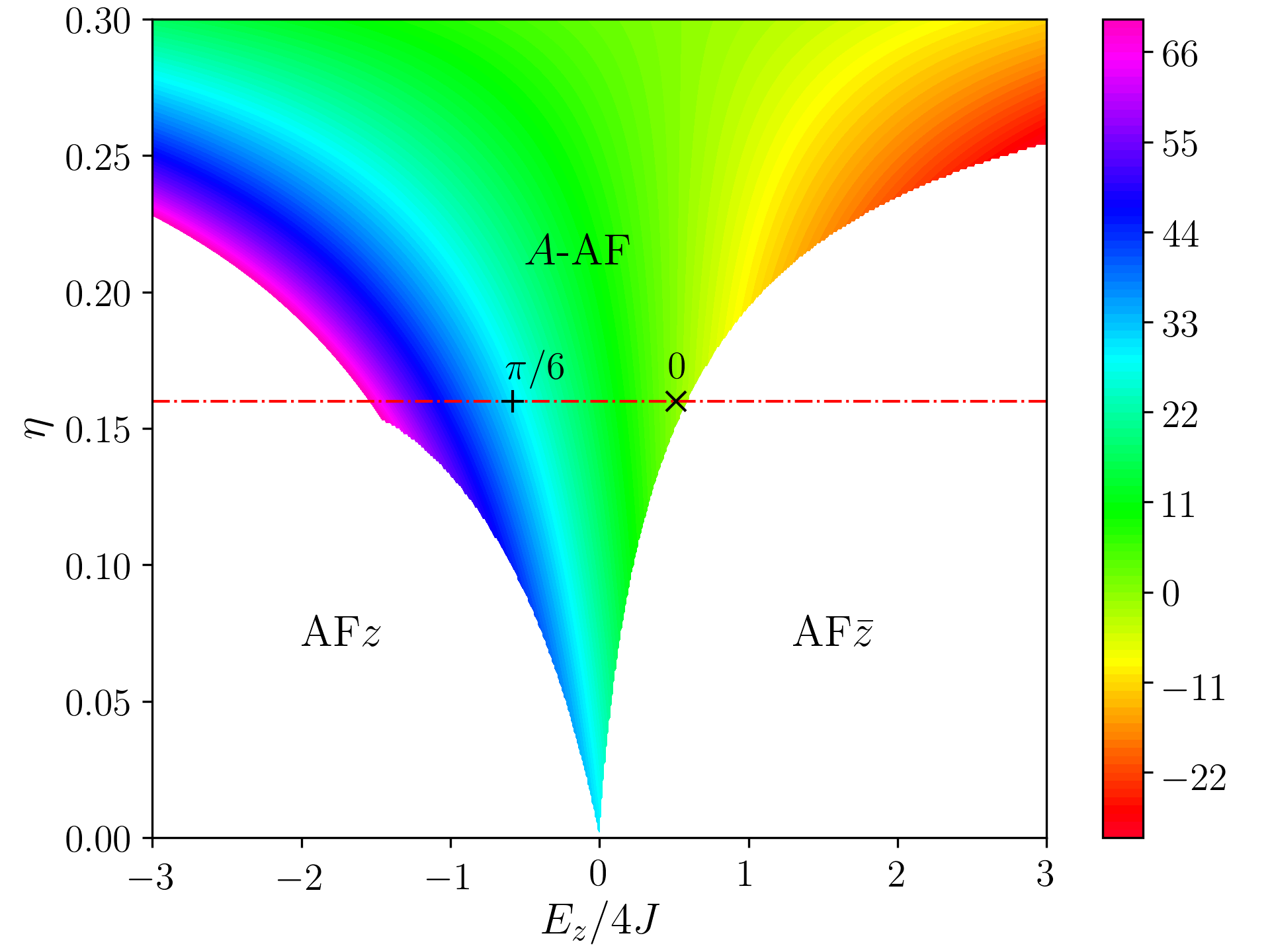}
\caption{The mean-field phase diagram of the 3D Kugel-Khomskii model.
We focus on the $A$-AF/$C$-AO spin-orbital order for which we determine 
the spectral function~\eqref{Aw} occurs between two AF phases with FO 
order (white areas), AF$z$ (left) and AF${\bar{z}}$ (right). 
The color scale indicates the detuning angle $\phi$ in degrees.
The values of $\phi=0$ and $\phi=\pi/6$, found at $\eta=0.16$
(red dashed line), used to investigate the spectral functions in the
present study, are indicated by $\times$ and +, respectively. Note that
a more complete mean-field phase diagram including possible phases
described by variational wave functions with short-range order was
presented before in Ref.~\cite{Brz13}.}
  \label{fig:phd}
\end{figure}

In this way we generate analytical EOMs that easily allow for exact
implementation of the local constraints (\emph{i.e.}, charge and bosons are
forbidden from being at the same site simply by removing from the
variational space the configurations which violate this constraint).
Once generated, the EOMs form an inhomogeneous system of linearly
coupled equations, which is solved numerically to yield all the Green's
functions, and in particular $G(\vect{k},\omega)$ from which we
determine the spectral function,
\begin{equation}
A(\vect{k},\omega)=-\frac{1}{\pi}\Im G(\vect{k},\omega).
\label{Aw}
\end{equation}
This quantity is directly measured through angle resolved photoemission
spectroscopy for LaMnO$_3$, or inverse photoemission for KCuF$_3$.

%\textcolor{blue}{
We shall be interested in the spectral function obtained for the 
$A$-AF/$C$-AO spin-orbital order phase where both magnon and orbiton 
excitations may couple to the moving charge. The mean field analysis 
of this phase includes the energy minimization to select the optimal 
value of the detuning angle $\phi$, as described in Appendix A. 
We investigate two ground states with $\phi=0$ and 
$\phi={\pi}/{6}$ found at 
$\eta=0.16$, shown by the respective symbols in Fig.~\ref{fig:phd},
and take $t\equiv 1.0$ as the energy unit.%} 

%\textcolor{blue}{
Our method, while highly accurate and versatile, does not come without 
its limitations. The most important stems from the very basis of the 
expansion, namely the cut-off criterion being implemented in real 
space. As a consequence, only local processes can be treated exactly, 
while other interactions have to be approximated in a way compatible 
with this methodology. As such, this method is especially well-suited 
to polaronic problems, where a charge couples to bosonic excitations 
either on-site or on the nearest-neighboring site, such as in this 
paper. The most common obstacle here is the treatment of quantum 
fluctuations, which are not tied to the itinerant charge and are 
therefore completely non-local. These are generally treated by being 
included only in the immediate neighborhood of the electron, the logic 
behind this being that only then will they affect the properties of 
the arising QP. This works as long as the classical ground state is 
not too different from the true quantum ground state, \emph{i.e.}, 
the classical state is a good starting point for the expansion. 
This would make our method tricky to use in 1D, but any higher 
dimensional problem is easily treatable. Another limitation comes from 
the use of real space Green's functions, which are hard to calculate 
already for a single electron. Treatment of multi-electron problems is 
an ongoing, highly challenging effort, although this is certainly true 
of all semi-analytical Green's function methods. Here we only focus on 
single-electron spectral functions, which are relevant for 
photoemission spectroscopies.%}

\begin{figure*}[t!]
  \includegraphics[width=0.49\textwidth]{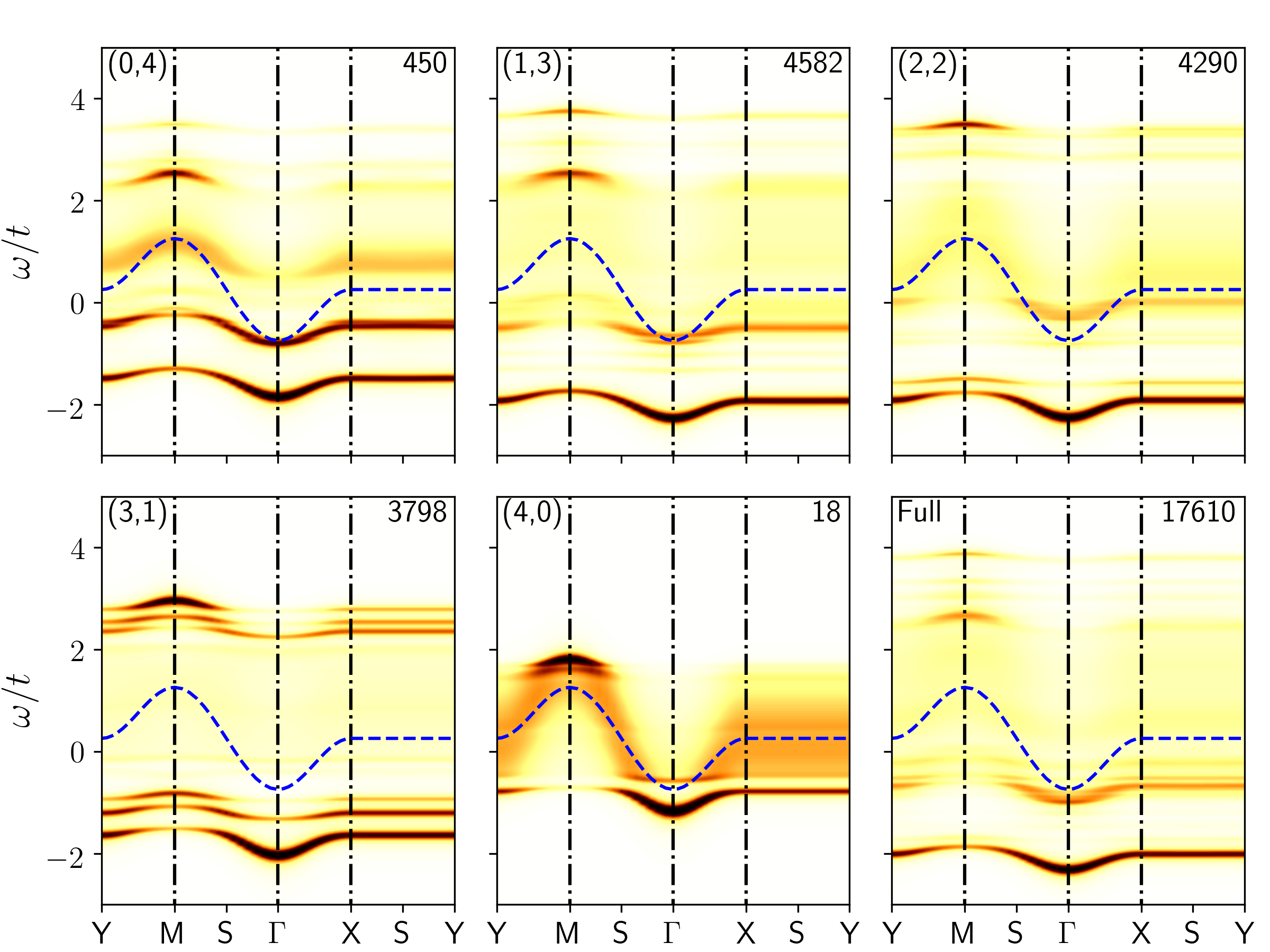}
  \hskip .01cm
  \includegraphics[width=0.49\textwidth]{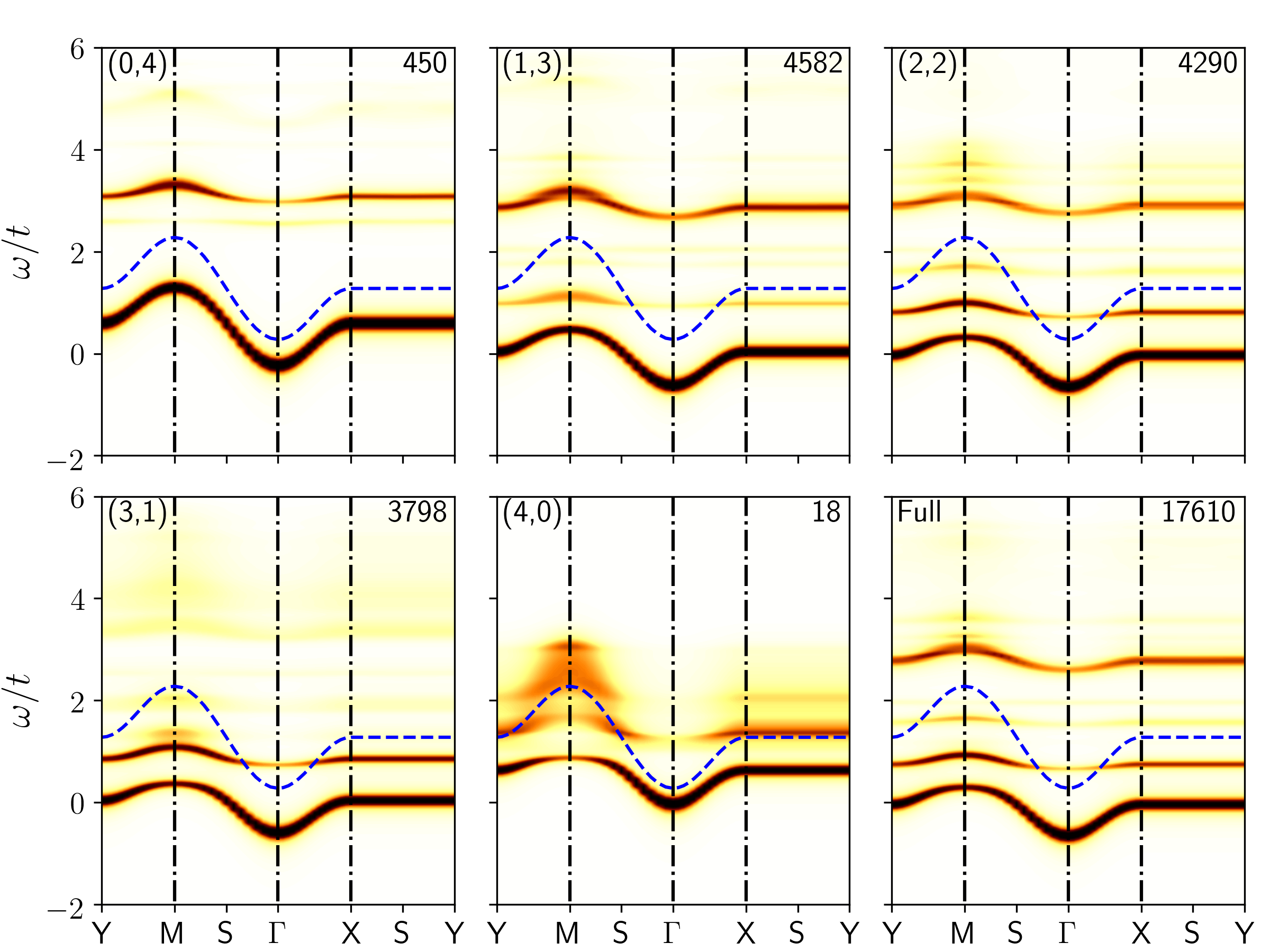}
\caption{The spectral functions $A(\vect{k},\omega)$ (shown by intensity
of brown/yellow color) in partial and full variational space for
$\phi=0$ and $J=0.1$ (left) and $J=0.5$ (right). The dashed blue line
indicates the free charge dispersion, $\epsilon_{\vect{k}\phi}$.
The numbers in the upper-left corner indicate the maximal number of
magnons and orbitons, respectively. The number in the upper-right
corner gives the size of the variational space.
The high-symmetry points are: $\Gamma=(0,0)$, $X=(\pi,0)$, $Y=(0,\pi)$,
$S=(\pi/2,\pi/2)$, and $M=(\pi,\pi)$. Parameter: $\eta=0.16$.}
  \label{fig:1}
\end{figure*}

\section{Results and discussion}
\label{sec:res}

We carry out the MA calculation in the variational space defined by
configurations with up to 4 bosons present. Because the calculation is
done for a 3D system with full treatment of the charge coupling to 
bosonic degrees of freedom, the branching factor for the EOMs is far 
too great to allow us to include more configurations. Nevertheless, 
based on our previous research within similar models~\cite{Bie16,Bie17}, 
we expect this choice to be sufficient for the ground state convergence 
to be satisfactory.

In order to distinguish the physical effects arising due to the
coupling to magnons and orbitons, we have performed the calculation not
only in the variational space with up to four bosons of any kind, but
also in subspaces where we further restrict the number of individual
bosonic flavors (\emph{e.g.}, up to three orbitons and up to one magnon).
This allows us, to some extent, to trace the evolution of the spectral
function depending on the bosonic content of the QP's cloud in its
ground state. By comparing these subspace projections to the full
calculation, we can infer which bosons dominate the QP dynamics.

The spectral functions~\eqref{Aw} were obtained for two representative
mixing angles $\phi$ with coexisting $A$-AF/$C$-AO spin-orbital order,
$\phi=0$ and $\phi=\pi/6$. 
They occur at finite Hund's exchange $\eta>0$ near the orbital 
degeneracy, $E_z\approx 0$. We have selected $\eta=0.16$ which is 
representative for the AO order in KCuF$_3$ considered here and close 
to what is reported in earlier studies 
\cite{Ole05,Brz13,Lie95,Kat04,Pav08,Leo10}. This value ensures that 
both the $\phi=0$ and $\phi=\pi/6$ $A$-AF/$C$-AO phases appear as the 
actual ground states within the range of variation of the crystal 
field~\cite{Brz13}.

The first \mbox{$A$-AF/$C$-AO} spin-orbital phase is obtained for 
$E_z>0$ close to the boundary between $A$-AF and AF$\bar{z}$ phases, see 
Fig.~\ref{fig:phd}. It is characterised by symmetric and antisymmetric 
linear combinations of the basis orbitals, 
$\{|\bar{z}\rangle,|z\rangle\}$; a finite value of $E_z>0$ is needed 
because of the spin order which is AF in the $(a,b)$ planes and FM 
along the $c$ axis. 
The second spin-orbital phase discussed below has the orbital angle 
$\phi={\pi}/{6}$ (\ref{phi}), which is obtained for $E_z<0$, see Fig. 
\ref{fig:phd}. It corresponds to the other extreme characterized by 
the external orbital field favoring the Kugel-Khomskii orbitals.

\begin{figure}[b!]
  \includegraphics[width=\columnwidth]{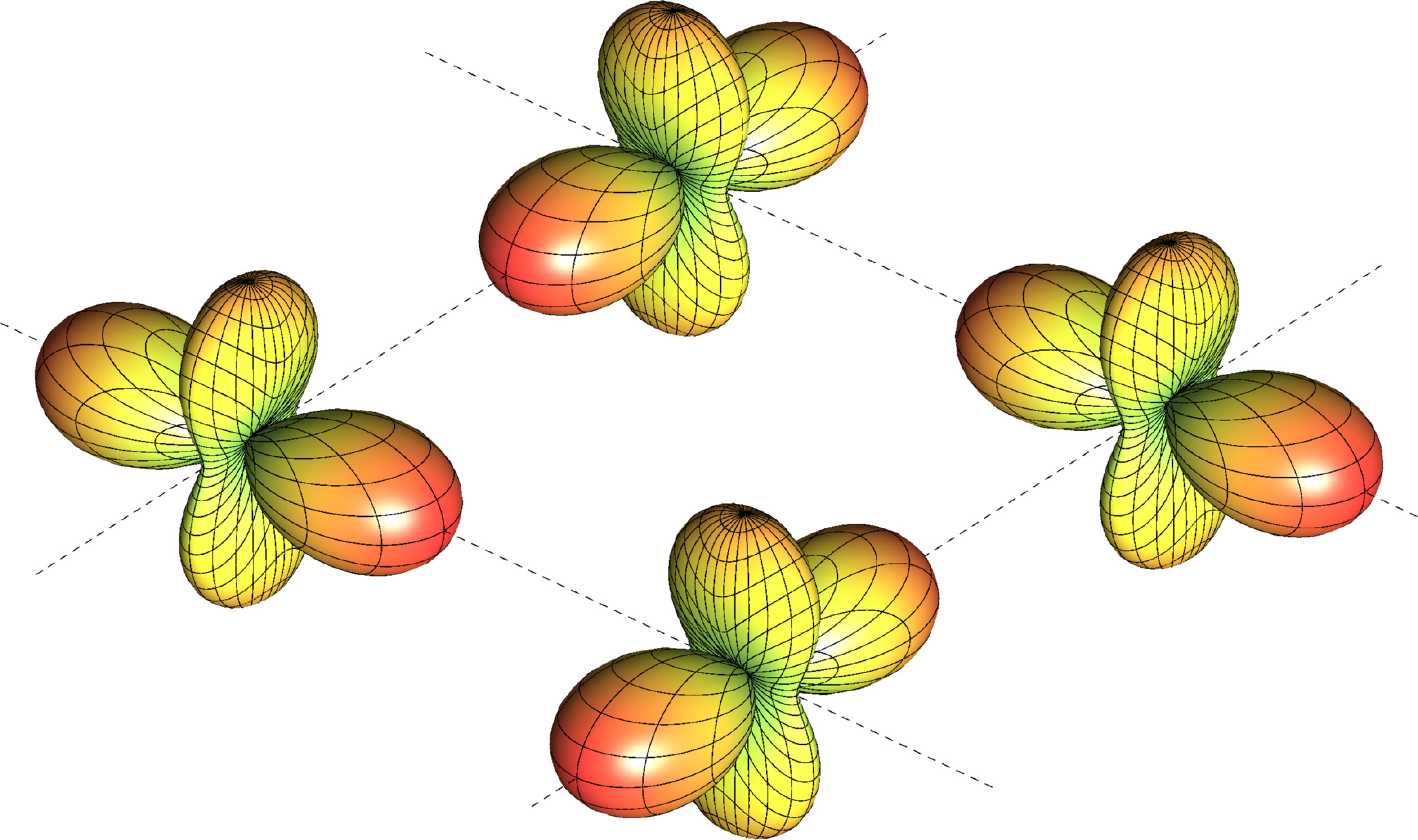}
  \caption{The in-plane orbital arrangement of the $\phi=0$ phase.}
  \label{fig:gs0}
\end{figure}

We start by analyzing the spectral functions for the $\phi=0$ phase in
%\textcolor{blue}{(shown in Fig.~\ref{fig:gs0})} in
the Ising limit, see Fig.~\ref{fig:1}. 
%\textcolor{blue}{
The occupied $|\pm\rangle$ orbitals (\ref{pm}) form an AO state shown
in Fig.~\ref{fig:gs0}.%} 
The Ising limit used here is defined by
neglecting both spin and orbital fluctuations, \emph{i.e.}, discarding
all terms containing operators other than $S_{}^z$ or $T_{}^z$. Note
that the spectral function density maps are presented in a nonlinear
$\propto\tanh$ scale which allows us to highlight the low amplitude
states that would otherwise not be visible. The results are shown for
two values of the superexchange constant, $J=0.1$ (canonical value,
note that the definition of \mbox{$J\equiv t^2/U$} does not include 
here the factor of 4, conventionally present in the standard $t$-$J$ 
model) and $J=0.5$ (weak interaction regime, 
%\textcolor{blue}{
this is not a physically relevant limit but it is useful for exploring 
the interdependence between orbitons and magnons in the system, and its
effect on the polaronic physics).

\begin{figure}[t!]
  \includegraphics[width=\columnwidth]{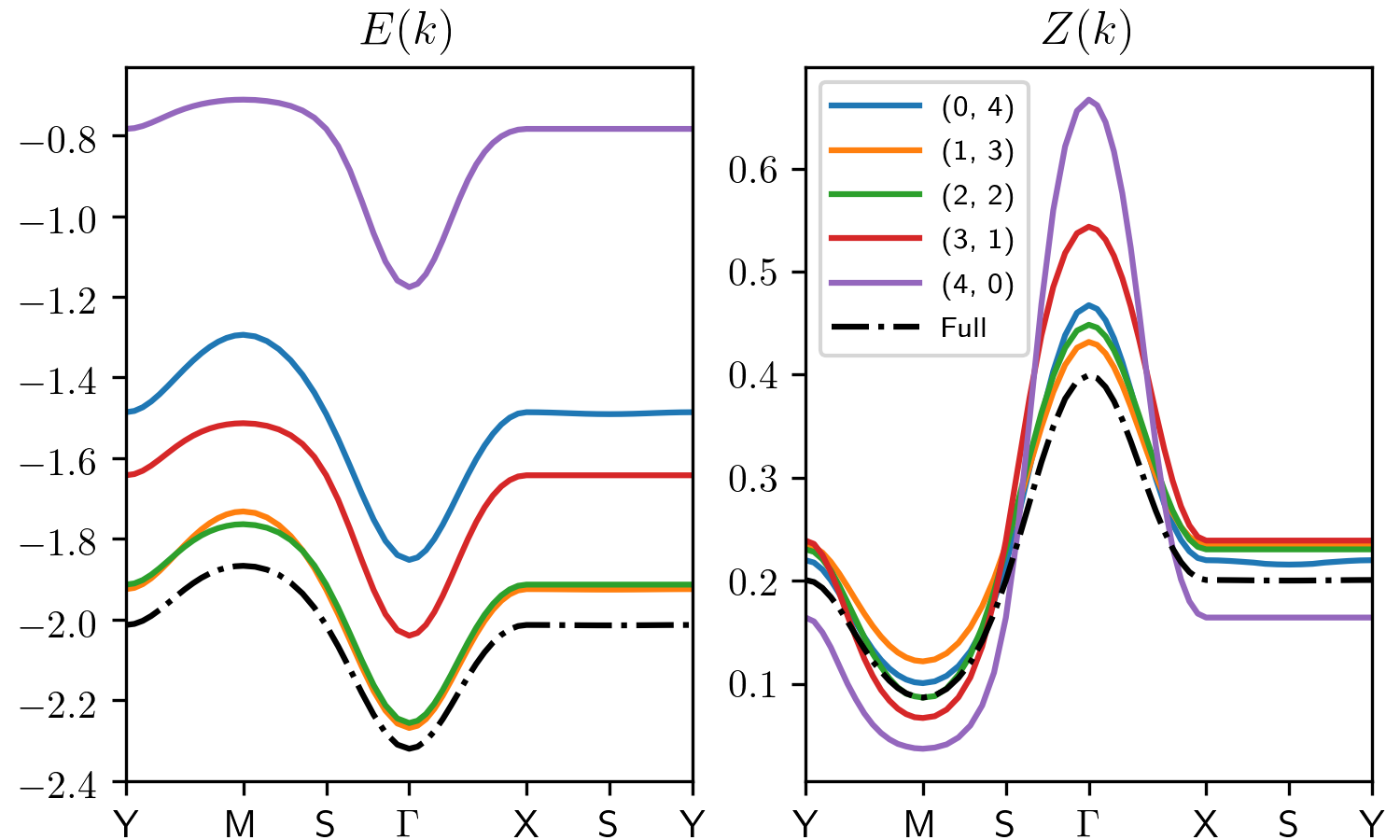}\\
  \includegraphics[width=\columnwidth]{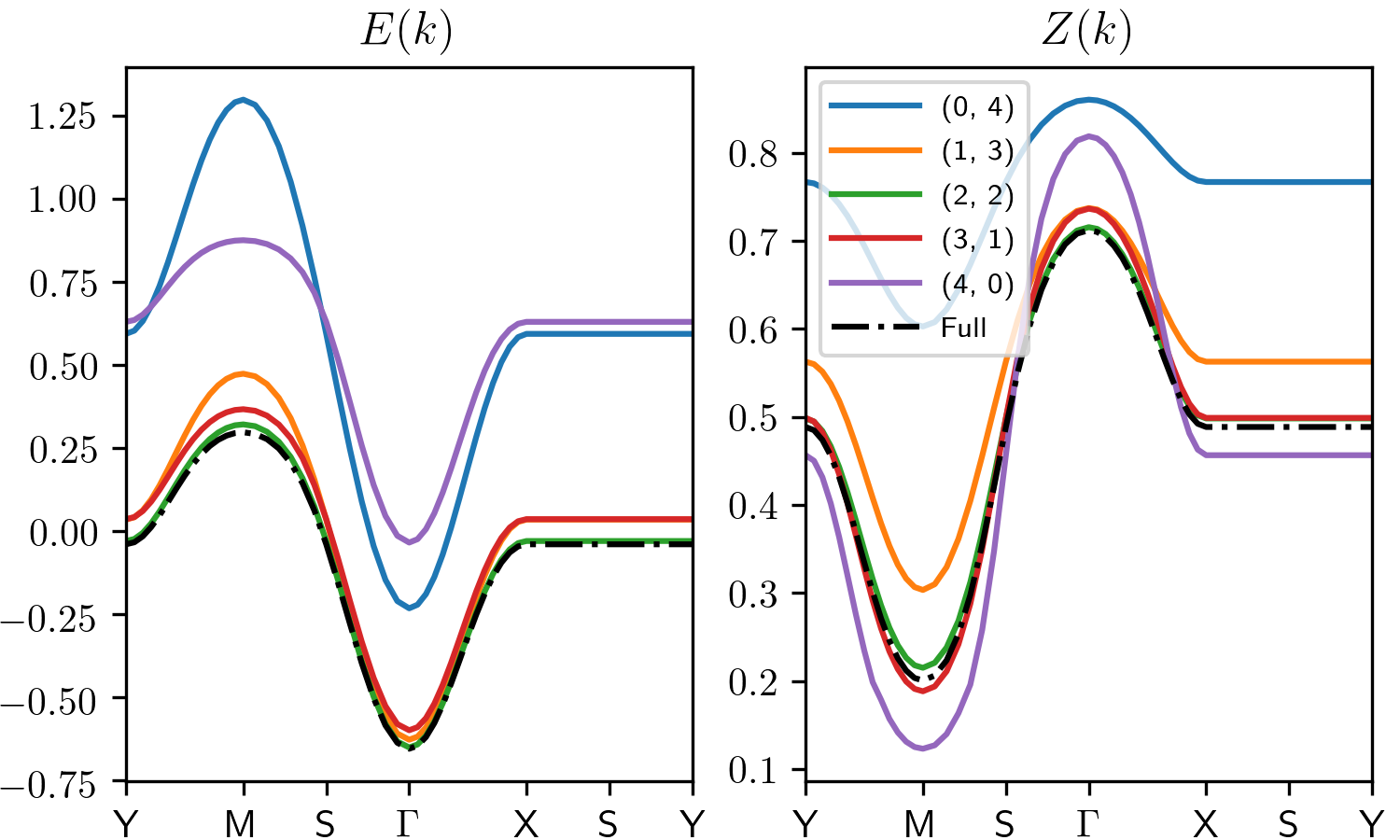}
\caption{The extracted QP ground state energies $E(\vect{k})$ (left) 
and spectral weights $Z(\vect{k})$ for the full fourth order expansion 
(right), and the respective subspace expansions for the exchange 
constants $J=0.1$ (upper panels) and $J=0.5$ (lower panels). 
Parameter: $\eta=0.16$.
Labeling conventions are the same as in Fig.~\ref{fig:1}.}
  \label{fig:2}
\end{figure}

Each panel in Fig.~\ref{fig:1} is marked in the upper-left corner with
the maximal number of magnons and orbitons, respectively, allowed in a
given subspace, and in the upper-right corner with the size of the
variational Hilbert space. The lower-right panel marked with the word
``Full'' presents the full expansion for up to 4 bosons (without further
specifying individual bosonic flavors). The dashed blue line indicates
the free charge dispersion $\epsilon_{\vect{k}\phi}$, and serves as a
reference energy for the QP state. As expected, the dressing with
bosons creates a QP which is energetically more stable than the free
particle, however this comes at the cost of an increased effective mass
and decreased mobility. Note that this is all consistent with standard
polaronic physics. The renormalization is
much smaller for the large $J$ limit. This can be easily understood
because the cost of creating any boson is proportional to~$J$, so the
bigger $J$ is, the more expensive it is to create a big bosonic cloud.
Thus, for large $J$ there will be fewer bosons in the cloud, resulting
in smaller renormalization of physical properties.

Remarkably, by comparing the full results against the partial results,
we can see that in the strong interaction case ($J=0.1$) the QP
behaves predominantly like in the orbiton rich cases (1,3) and (2,2).
To highlight this effect we extract the ground state energy and
spectral weight for all these solutions and plot them against each
other, see Fig.~\ref{fig:2}. As is evident, the full solution tends
to include more orbitons and fewer magnons. Having said that though,
a cloud consisting of only orbitons would not be sufficient to achieve
the optimal QP energy, either. Thus, we can already see that this is an
intrinsically spin-orbital system, where the interaction of \emph{all}
degrees of freedom (charge, spin, and orbital) is crucial to achieve
the complete understanding of underlying physics.

\begin{figure}[t!]
  \includegraphics[width=\columnwidth]{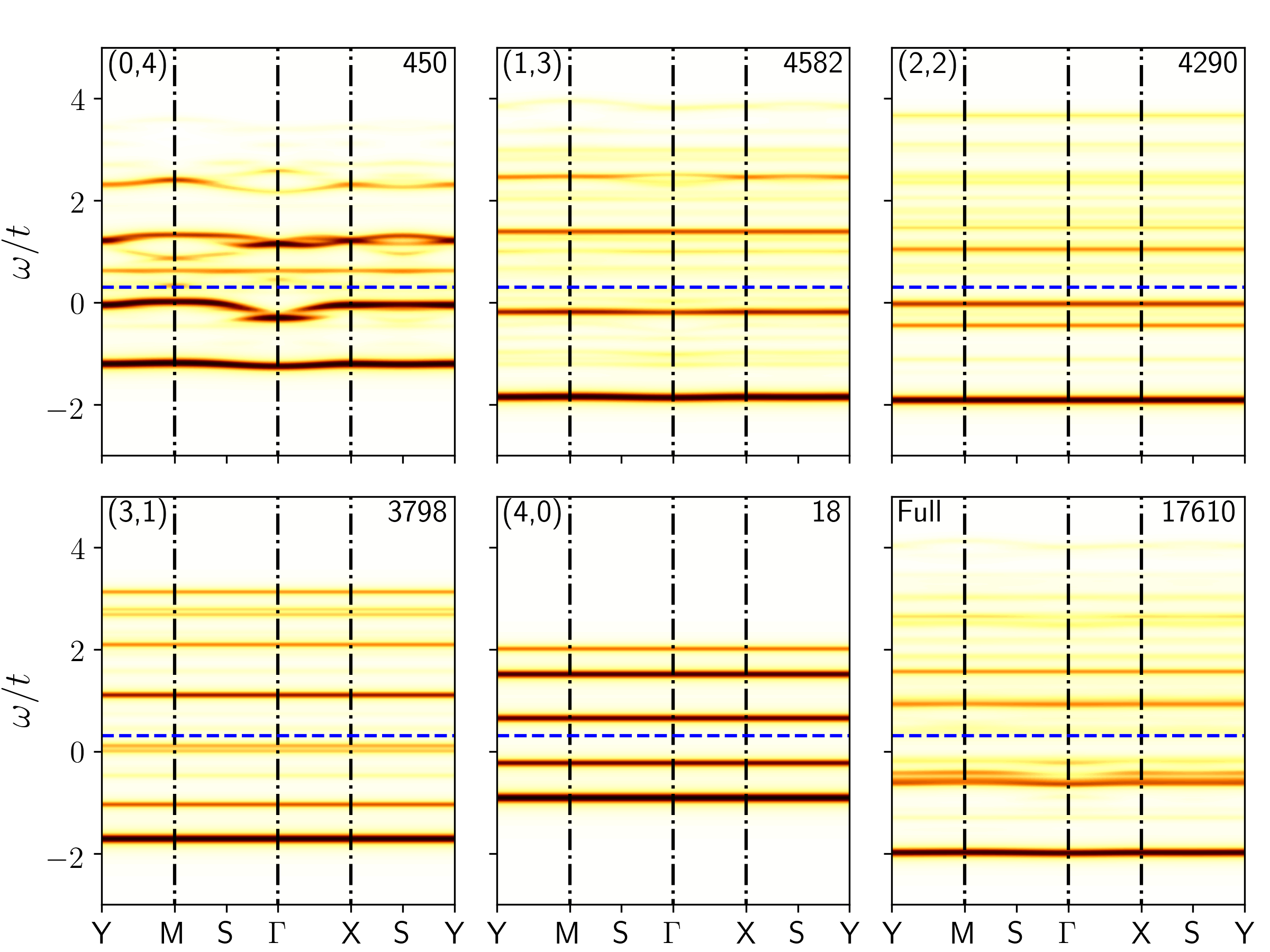}
\caption{The full and partial spectral functions $A(\vect{k},\omega)$
for the $\phi=\pi/6$ phase. Parameters: $J=0.1$ and $\eta=0.16$.
Notation and conventions are the same as in Fig. 2. }
  \label{fig:3}
\end{figure}

Even more interestingly, if we now make the same comparison for the
weak interaction limit ($J=0.5$), we see that this time the QP band
behaves most like the magnon-rich solutions (3,1) and (2,2). This
suggests a crossover, controlled by the exchange parameter $J$, between
orbiton-rich and magnon-rich QP clouds. This happens because magnons
have lower energy and are cheaper to create than orbitons. In the large
$J$ limit, only very few bosons are created and they are more likely to
be magnons, which therefore dominate the dynamics of the resulting QP.
In contrast, for small $J$ all bosons are cheap(er) and orbitons
dominate by means of geometric effects, \emph{i.e.}, the fact that the
charge can couple to them by moving in any of the three principal cubic
directions, in contrast to magnons which couple only when the particle
moves along the \emph{single} AF $c$ direction~\cite{Bie17}.

\begin{figure}[t!]
  \includegraphics[width=\columnwidth]{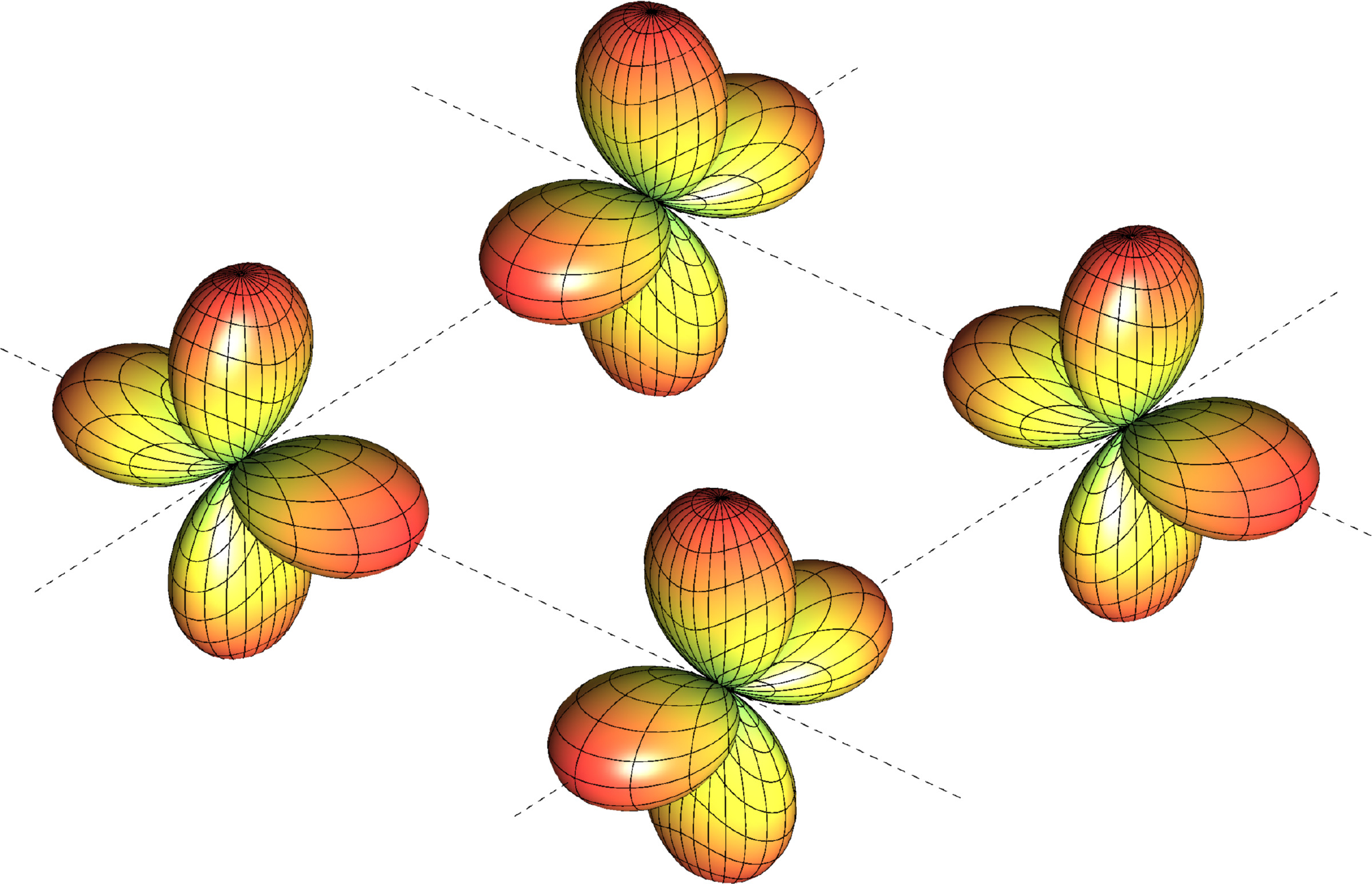}
  \caption{The in-plane orbital arrangement of the $\phi=\pi/6$ 
  Kugel-Khomskii phase.}
  \label{fig:gs30}
\end{figure}

Figure~\ref{fig:3} shows the spectral functions for  $\phi=\pi/6$ with
$J=0.1$. 
%\textcolor{blue}{
The orbital order itself is depicted in Fig.~\ref{fig:gs30}.%}
The first striking observation is that the bands show hardly
any dispersion at all, except for the purely orbitonic solution (0,4).
This is easily understood if we look at the free charge dispersion 
$\epsilon_{\vect{k}\phi}$, which vanishes for $\phi=\pi/6$, as 
illustrated by the flat dashed blue reference line in Fig.~\ref{fig:3}. 
In other words,the unrenormalized particle is completely localized, and 
the coupling to bosons does not change that in any substantial way. The 
tiny dispersion observed in the orbitonic solution is due to Trugman
loops~\cite{Tru88}, which require a 2D AO order, just like we
have in this system, and the existence of at least three-boson clouds,
hence its appearance in the purely orbitonic solution. In fact, a very
tiny dispersion can also be seen in the (1,3) panel, however there the
interference between orbitons and magnons clearly suppresses the
Trugman processes~\cite{Tru88}, again underlining the crucial role of
orbiton-magnon interplay in the physics of these systems.

The lack of dispersion in this orbital phase is a straightforward
consequence of a special symmetry of the orbital order in the
Kugel-Khomskii state. Namely, as evident from Fig.~\ref{fig:gs30}, 
the $\phi=\pi/6$ detuning corresponds to the occupation of AO
$y^2-z^2/z^2-x^2$, so the hopping process would require the charge 
to move from a lobe of one such orbital to the nodal point of the 
neighboring orbital, which is forbidden by symmetry 
%\textcolor{blue}{
of the wave function.%}

The results presented thus far point to an interesting experimental
possibility. Namely, the orbital order should be discernible from a
spectral experiment: the flatter the QP band, the closer the occupied
orbitals should be to the $\phi=\pi/6$ phase. Naturally, determining
the exact phase might not be simple, however, verifying the validity
of the $\phi=\pi/6$ case to which most local density approximation
(LDA) studies seem to
point~\cite{Kat04,Pav08,Leo10,Bin04,Pav10} should be possible owing
to the dispersionless character of this phase.
Having said that, the issue of an insulating sample and thus strong
charging during an angle resolved photoemission spectroscopy (ARPES)
experiment might pose a barrier even to this verification.

\begin{figure}[t!]
  \includegraphics[width=\columnwidth]{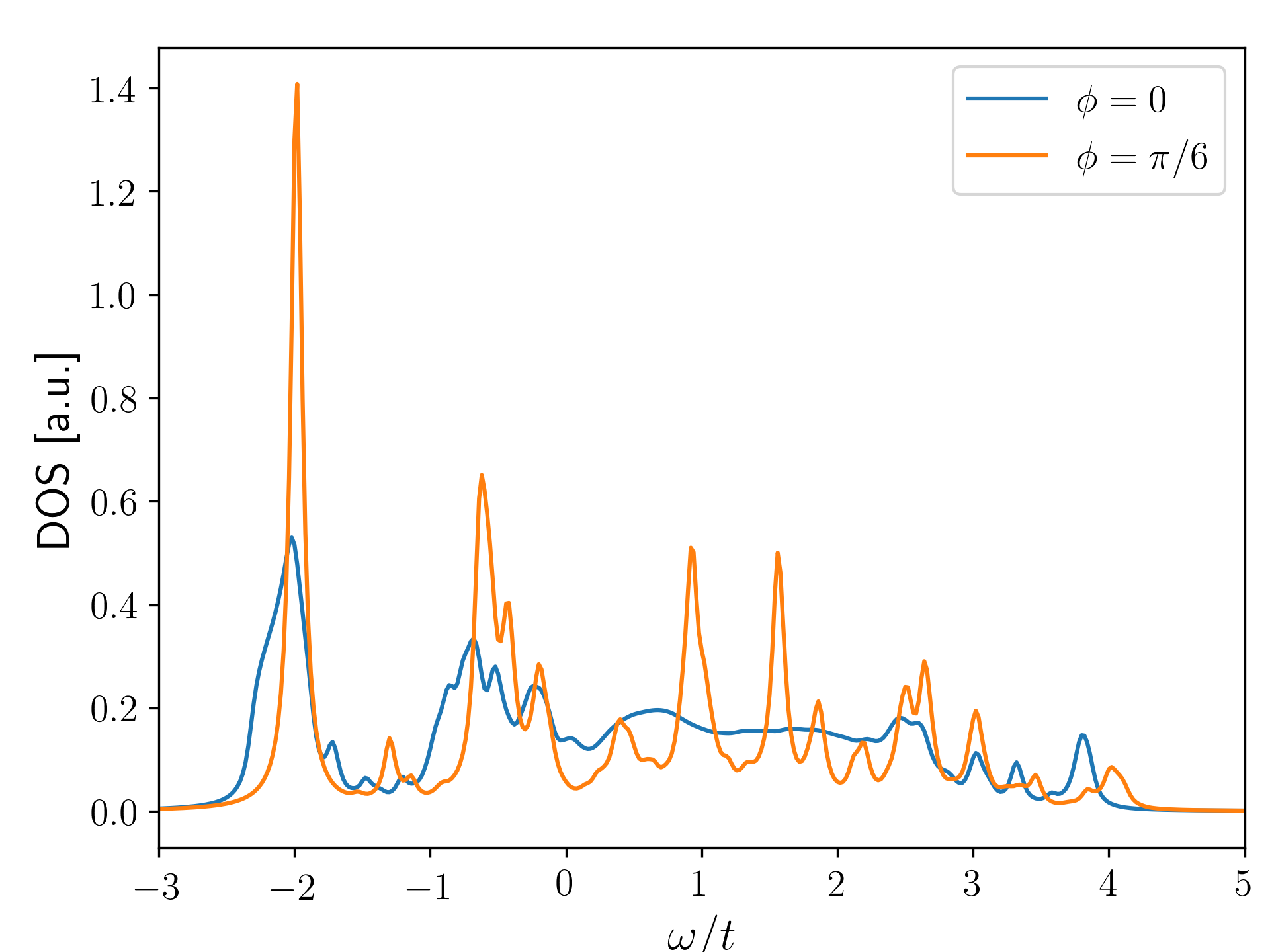}
\caption{Comparison of the density of states for the two major orbital
phases discussed in this paper, $\phi=0$ and $\phi=\pi/6$;
Parameters: $J=0.1$ and $\eta=0.16$.}
  \label{fig:4}
\end{figure}

There is another possibility, however, owing to the QP mass
renormalization. Going back to Fig.~\ref{fig:1} and comparing the QP
{\em vs.} the free charge dispersion, we see that not only is there a
difference in bandwidth between the two cases, but also the symmetry
between the $\Gamma$ and $M$ points is significantly suppressed, with
the QP band at the $M$ point being much flatter and having a greatly
reduced spectral weight. If we would now integrate the spectrum to
produce the density of states (DOS) for this system, we would see that
the QP DOS for a dispersive phase should be highly asymmetric, whereas
the dispersionless phase should be characterized by a sharp and
completely symmetric QP DOS, as verified in Fig.~\ref{fig:4}. Thus, the
orbital phase could be inferred, even if only approximately, from the
shape and asymmetry of the QP DOS. In turn, the DOS can be obtained
from a scanning tunneling microscope experiment for which sample
charging might be less problematic.

In all of the above we have assumed an Ising interaction, or to put it
differently, that the effect of quantum fluctuations is negligible.
This is a reasonable assumption because fluctuations generally are less
important in higher dimensions and here we are dealing with an
ostensibly 3D system. Having said that, however, the AO order can at
the same time be thought as 2D and the AF order as 1D. While it has
been established that the role of fluctuations for $e_{g}$ orbital
pseudospin in a 2D planar subsystem is indeed negligible~\cite{Bie16},
the same assumption seems less justified for magnetic excitations.
Apart from the dimensionality of the corresponding order, another
argument is that the relative lack of importance of orbitonic
fluctuations comes from the fact that the orbiton spectrum is gapped,
which is not the case for magnons. This is why it is reasonable to
neglect the orbital fluctuations while including magnetic
fluctuations, in order to explicitly establish whether they are
relevant or not.

Fortunately, magnetic fluctuations may be fairly easily included within
MA by allowing arbitrary fluctuations but only in the vicinity of the
propagating charge (the variational space cutoff is controlled with
exactly the same cloud spatial criteria as before), since these are
the only ones which will affect the QP dynamics. Fluctuations occurring
far from the charge will instead only affect the nature of undoped
regions far from the particle, affecting the overall energy. However,
as long as the classical ground state is close enough to the true
quantum state realized for a given set of parameters, this would only
be reflected by a constant shift of the entire spectrum, which is not
a physically significant effect.

\begin{figure}[t!]
  \includegraphics[width=\columnwidth]{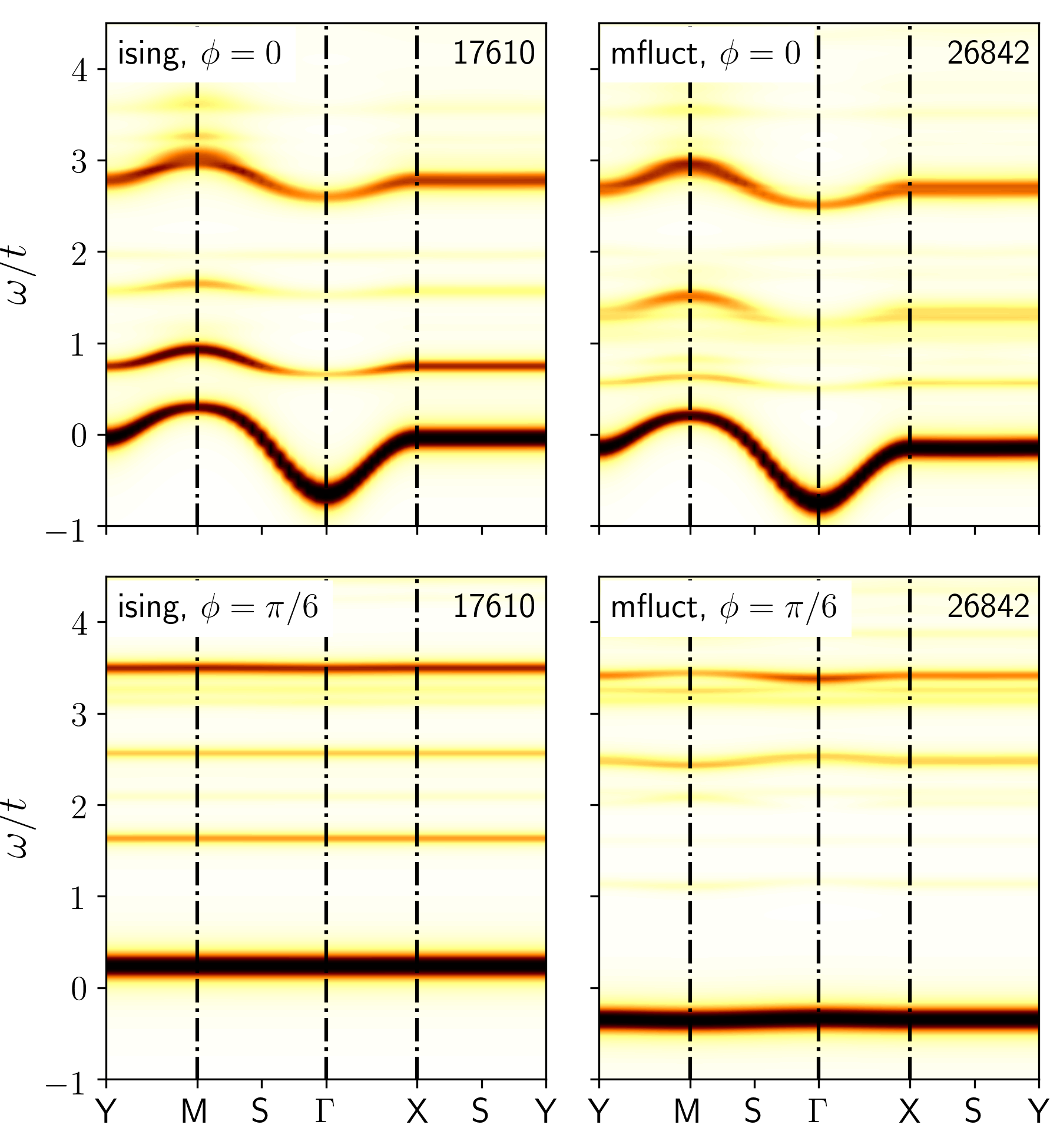}
\caption{Comparison of the spectral functions in the Ising
approximation (ising) and the one including full magnetic fluctuations
(mfluct), and for the two orbital phases, $\phi=0$ and $\phi=\pi/6$.
To highlight the effect we take the weak interaction regime
\mbox{$J=0.5$}. Notation and conventions are the same as in Fig. 2.
Parameter: $\eta=0.16$.}
  \label{fig:5}
\end{figure}

To illustrate the role of fluctuations, we present a comparison between
the Ising solution and the one including local fluctuations for both
angles, $\phi=0$ and $\phi=\pi/6$, see Fig.~\ref{fig:5}. As their
effect proves to be rather elusive, we focus on the weak interaction
limit $J=0.5$, where the changes can be more readily observed. One
immediately sees the huge difference in the size of the variational
spaces, even though only magnetic fluctuations (albeit in all three
cubic directions) are considered here. This, however, has surprisingly
little overall effect on the QP dispersion. While their effect seems
more considerable for the excited states, these are likely not fully
converged anyway, so that part of the spectra is not sufficiently
reliable for comparison. A tiny dispersive effect can also be observed,
most readily visible in the $\phi=\pi/6$ phase, however it is much too
small to be of any practical importance.

There is however one interesting feature, namely the pure gain in
energy experienced by the QP ground state, indicative of a stronger
binding of the QP, which however does not affect its dynamical
properties. In particular, while the gain for the $\phi=0$ phase is
relatively small, the one observed for $\phi=\pi/6$ is considerable.
This difference could be indicative of a subtle quantum effect arising
from the spin-orbital coupling in the system. Clearly, the importance
of the magnetic fluctuations strongly depends on the orbital phase in
the system, and in particular, the fluctuations in the $\phi=\pi/6$
phase grow particularly strong. This indicates that the magnetic order
becomes less classical in character, which is likely caused by the
system decoupling into 1D AF chains. There is ample evidence of the
actual KCuF$_3$ exhibiting a 1D quantum AF character~\cite{Lak05}, so
this would seem to point to the actual orbital phase in that system
being close to $\phi=\pi/6$, something that was long proposed based on
electronic structure calculations using LDA.
Here we were able to arrive at similar conclusions
through indirect means and by a completely different methodology.

\section{Summary and Conclusions}
\label{sec:summa}

We have developed an effective spin-orbital superexchange model for an
$e_g^3$ system, and computed the single-polaron spectrum resulting when
a single charge is doped in the system by using the semi-analytical,
variational momentum average method for calculating Green's functions.
This allowed us not only to obtain the relevant spectral functions, but
also to gain insight into the nature of the magnetic and orbital order
in the system. Thus we were able to demonstrate a number of subtle
quantum effects arising from the interaction between the charge,
orbital, and magnetic degrees of freedom.

One such effects is the change of the character of the polaronic 
quasiparticle cloud from being dominated by orbitons to being
dominated by magnons; this is controlled by the strength of the
superexchange interaction $J$. This behavior, although only a
theoretical prediction due to the impossibility of experimentally
tuning the parameter $J$ in such a wide range of values, nonetheless
points towards a strong interplay between orbital and magnetic degrees
of freedom in this model. It should come as no surprise, then, that
their intermingling should also crop up in other properties of the
system, some of which might be more readily accessible to experiment.

One possible experimental consequence lies in the quasiparticle 
dispersion being strongly dependent on the orbital order in the system 
which, coupled with the polaronic suppression of the symmetry between 
the $\Gamma$ and $M$ points of the Brillouin zone, suggests that the
quasiparticle density of states should be particularly sensitive to
the orbital order in these systems. In turn, this would point towards
Scanning Tunneling Microscopy as a promising tool for an experimental
probe of the orbital order type. We thus propose that the orbital order
could be inferred by investigating the amplitude to width ratio and the
asymmetry of the density of states peaks.

Finally, we point out that the orbital order around the detuning angle
$\phi=\pi/6$ seems to drive the magnetic system closer towards the 1D
AF chain. Indeed, the already available results of neutron scattering
experiments~\cite{Lak05} demonstrate a nearly ideal 1D spin liquid
behavior. We suggest that this is a strong indication that the orbital
order in KCuF$_{3}$ is likely to be close to $\phi=\pi/6$.

\begin{acknowledgments}
K. B. and A. M. O. kindly acknowledge support by UBC Stewart Blusson
Quantum Matter Institute (SBQMI), by Natural Sciences and Engineering
Research Council of Canada (NSERC), and by
Narodowe Centrum Nauki (NCN, Poland) under Projects
Nos.~2016/23/B/ST3/00839 and 2015/16/T/ST3/00503.
\mbox{M. B. acknowledges support from SBQMI and NSERC.}
A.~M.~Ole\'s is grateful for the Alexander von Humboldt 
Foundation Fellowship (Humboldt-Forschungspreis).
\end{acknowledgments}

\appendix

%\section{Model derivation}
%\label{sec:deriv}

\section{The mean field ground state}
\label{sec:mfa}

In this Section we provide the more technical details concerning the 
derivation of the effective polaronic spin-orbital model that is the 
basis of our calculation.
In order to find the classical orbital ground state, we parametrize the
orbital basis in terms of a standard rotation of the $\{\bar{z},z\}$
basis. However, since the reference, field-free order is composed of
alternating $\ket{\pm}$ states~\eqref{pm}, the rotation is most
conveniently parametrized with an angle $\pm(\pi/2+\phi)$, where the
sign depends on the orbital sublattice, 
%with $\phi=0$ corresponding to the $\pm$ reference basis. 
%%AM
\begin{eqnarray}
\label{phi}
\ket{\phi_A}&=&
 \cos\left(\frac{\pi}{4}+\frac{\phi}{2}\right)\ket{\bar{z}}
+\sin\left(\frac{\pi}{4}+\frac{\phi}{2}\right)\ket{z},     \nonumber \\
\ket{\phi_B}&=&
 \cos\left(\frac{\pi}{4}+\frac{\phi}{2}\right)\ket{\bar{z}}
-\sin\left(\frac{\pi}{4}+\frac{\phi}{2}\right)\ket{z}.
\end{eqnarray}
This choice also serves to transform the 
underlying AO order to an FO order, effectively eliminating the 
bipartite division of the lattice in the reference state.
%\textcolor{blue}{
It should be stressed that this operation does not affect the actual 
ground state, it merely changes its representation to one that is more 
convenient---it spares us the trouble of distinguishing between bosons 
on different sublattices (\emph{cf.}~Ref.~\onlinecite{Mar91}).%}
Now $\phi$ will indicate a detuning from the field-free
orbital order, and the angle between the \emph{occupied} orbitals on
the two sublattices will be $2\phi$ (\emph{i.e.}, in general, the bases on
different sublattices will not be mutually orthogonal).

To find the relation between the orbital field $E_{z}$ and the detuning
angle $\phi$ in Eqs. (\ref{phi}), we start by writing out the 
superexchange Hamiltonian in the new basis:
\begin{widetext}
\begin{subequations}
  \label{eq:Hphi}
\begin{align}
  \label{eq:Hphia}
  \mathcal{H}_{\mathrm{I}}^{\perp} &=
   J\sum_{\mean{ij}\perp c}
  \left(A_{\eta}\vect{S}_{i}\cdot \vect{S}_{j}+\tfrac{1}{4}B_{\eta}\right)
  \left\{
    \tfrac{2C_{\eta}}{A_{\eta}}-1-(2\cos2\phi+1)\,T_{i}^{z}T_{j}^{z}
    -(2\cos2\phi-1)\,T_{i}^{x}T_{j}^{x}\right.\nonumber\\
     &\qquad\left. +2 e^{i\vect{Q}\vect{R}_{i}}\sin2\phi\, 
     (T_{i}^{z}T_{j}^{x}-T_{i}^{x}T_{j}^{z})
    \mp\sqrt{3}\,(T_{i}^{x}T_{j}^{z}+T_{i}^{z}T_{j}^{x})
  \right\},\\
  \label{eq:Hphib}
  \mathcal{H}_{\mathrm{I}}^{\parallel} &= 2J\sum_{\mean{ij}\parallel c}
  \left(A_{\eta}\vect{S}_{i}\cdot \vect{S}_{j}+\tfrac{1}{4}B_{\eta}\right)
  \left\{
    \tfrac{C_{\eta}}{A_{\eta}}- \tfrac12
    +2\sin^{2}\phi\,T_{i}^{z}T_{j}^{z}
    +2\cos^{2}\phi\,T_{i}^{x}T_{j}^{x}
    -  e^{i\vect{Q}\vect{R}_{i}}
    \sin2\phi\,(T_{i}^{x}T_{j}^{z}+T_{i}^{z}T_{j}^{x})
  \right\},\\
  \label{eq:Hphic}
  \mathcal{H}_{\mathrm{II}}^{\perp} &= -JC_{\eta}\sum_{\mean{ij}\perp c} 
  \left(\vect{S}_{i}\cdot \vect{S}_{j}-\tfrac{1}{4}\right)
  \left\{\left[
\sin\phi\,(T_{i}^{z}+T_{j}^{z})\mp\sqrt{3}e^{i\vect{Q}\vect{R}_{i}}\cos\phi\right]
    (T_{i}^{z}-T_{j}^{z})\right.\nonumber\\
   &\qquad\left.-e^{i\vect{Q}\vect{R}_{i}}\cos\phi\,(T_{i}^{x}-T_{j}^{x})
    \mp\sqrt{3}\sin\phi\,(T_{i}^{x}+T_{j}^{x})
  \right\},\\
  \label{eq:Hphid}
  \mathcal{H}_{\mathrm{II}}^{\parallel} &= 2JC_{\eta}\sum_{\mean{ij}\parallel c}
\left(\vect{S}_{i}\cdot \vect{S}_{j}-\tfrac{1}{4}+\tfrac{E_{z}}{4JC_{\eta}}\right)
  \left[
    \sin\phi\,(T_{i}^{z}+T_{j}^{z})
    -e^{i\vect{Q}\vect{R}_{i}}\cos\phi\,(T_{i}^{x}+T_{j}^{x})
  \right],
\end{align}
\end{subequations}
\end{widetext}
where the last term incorporates the orbital field $\mathcal{H}_{z}$. 
Here, $\vect{Q}=(\pi,\pi,0)$ is the ordering vector for the $C$-AO 
state, and the resulting phase factor encodes the alternating nature of 
the orbital order. The symbol $\perp/\parallel$ refers to the cubic 
directions with respect to the $c$-axis. Note that the various 
superexchange terms of Eq.~\eqref{eq:Hexch} have been split into terms 
quadratic in $\{T_i^z\}$ operators ($\mathcal{H}_{\mathrm{I}}$) and 
linear in $\{T_i^z\}$ operators ($\mathcal{H}_{\mathrm{II}}$). 
The Hund's exchange (\ref{eta}) is now encoded in the three prefactors 
(if $\eta=0$, one finds $A_0=B_0=C_0=1$):
\begin{subequations}
\begin{gather}
  \label{eq:Hund}
  A_{\eta} = \frac{1-\eta}{(1+\eta)(1-3\eta)},\\
  B_{\eta} = \frac{1+3\eta}{(1+\eta)(1-3\eta)},\\
  C_{\eta} = \frac{1}{1-\eta^{2}},
\end{gather}
\end{subequations}
which themselves result from various combinations of the $r_{i}$
multiplet parameters listed above. Note that the exchange Hamiltonian
has also been shifted in energy so that the Ising energy for the ground
state of the system is set to zero. This is done merely for reasons of
convenience, so that the excitation energies are easier to track once
we start considering excitations in the system.

Next we evaluate the mean field energy assuming the classical ground
state to be \textit{A}-AF/\textit{C}-AO, as is known to be the case in
KCuF$_{3}$. We find:
\begin{multline}
  \label{eq:mf}
  E_{\mathrm{MF}} = \frac{1}{4}J(B_{\eta}-A_{\eta})\sin^{2}\phi
  -\frac{1}{8}J(A_{\eta}+B_{\eta})(2\cos2\phi+1)\\
  -J\left(C_{\eta}-\frac{E_{z}}{2J}\right)\sin\phi.
\end{multline}
This expression is then minimized with respect to the detuning angle
$\phi$, yielding the relation
\begin{equation}
  \label{eq:ez}
E_{z} = J\left[2C_{\eta}-\left(A_{\eta}+3B_{\eta}\right)\sin\phi\right].
\end{equation}
This identity can now be used to eliminate $E_z$ from the Hamiltonian
by replacing it with the detuning angle $\phi$. Note that if we now set
$\phi=0$, we will, seemingly paradoxically, get $E_z=2JC_{\eta}$, \emph{i.e.},
a finite orbital field corresponding to the field-free case. This is
due to the fact that the superexchange Hamiltonian already includes
terms linear in pseudospin operators which behave like an orbital 
field, and the external field works to compensate these terms. In other 
words, the exchange Hamiltonian breaks cubic symmetry by itself, and 
the above estimate is the orbital field needed to restore it. On the 
other hand, the case $E_z=0$ corresponds to $\phi=\pi/6$ in the 2D
orbital model~\cite{Cza17}, which is the Kugel-Khomskii state composed
of alternating $y^2-z^2/z^2-x^2$ states. These two limits are commonly
cited as the extreme possibilities for the orbital order in this system.
The actual orbital order realized in the system will be bounded by
these two extremes, and in fact could be, to some extent, tuned by 
means of an axial pressure $\propto E_z$ applied along the $c$ axis.

%%AM
%\subsection{The single particle Hamiltonian}
\section{Fermion-boson polaronic model}
\label{sec:flu}

To go beyond the mean field ground state, we derive the effective 
%%AM
%Hamiltonian governing the physics of a single particle doped into the 
%system. 
Hamiltonian transforming the physics of a single charge doped into a
spin-orbital model to a fermion-boson many-body problem. To that end, 
we transform the kinetic Hamiltonian to the same basis as the one 
considered in the last Section, so that the entire 
model is expressed in compatible representations. This leads to
\begin{subequations}
  \label{eq:Htso3}
\begin{align}
  \label{eq:Ht3perp}
%  \begin{split}
  \mathcal{H}_{t}^{\perp} &= -\frac{t}{4}\sum_{\mean{ij}\perp c,\sigma}
  \left\{\left[
    (1-2\sin\phi)d_{i\sigma0}^{\dag} d_{j\sigma0}^{}\right.\right.\nonumber\\
   &\qquad-2 e^{i\vect{Q}\vect{R}_{i}}\cos\phi\,
   (d_{i\sigma0}^{\dag} d_{j\sigma1}^{}-d_{i\sigma1}^{\dag}d_{j\sigma0}^{})
\nonumber\\
   &\qquad \mp\sqrt{3}\,(d_{i\sigma0}^{\dag} d_{j\sigma1}^{} 
             +d_{i\sigma1}^{\dag} d_{j\sigma0}^{})\nonumber\\
   &\qquad\left.\left. -(1+2\sin\phi)d_{i\sigma1}^{\dag} d_{j\sigma1}^{}
  \right] +\mathrm{H.c.}\right\}\,,
  % \end{split}
  \\
  \label{eq:Ht3para}
%  \begin{split}
  \mathcal{H}_{t}^{\parallel} &= -\frac{t}{2}\sum_{\mean{ij}\parallel c,\sigma}
  \left\{\left[
    (1+\sin\phi) d_{i\sigma0}^{\dag} d_{j\bar{\sigma}0}^{}\right.\right.\nonumber\\
  &\qquad -e^{i\vect{Q}\vect{R}_{i}}\cos\phi\,
  (d_{i\sigma0}^{\dag}d_{j\bar{\sigma}1}^{}+d_{i\sigma1}^{\dag} d_{j\bar{\sigma}0}^{})
\nonumber\\
  &\qquad\left.\left. -(1-\sin\phi)\,d_{i\sigma1}^{\dag} d_{j\bar{\sigma}1}^{}
  \right]+\mathrm{H.c.}\right\}\,,
%\end{split}
\end{align}
\end{subequations}
where the 0 (1) indices denote the ground (excited) orbital states,
respectively.

Finally, following Mart\'inez and Horsch~\cite{Mar91}, we represent the
spin and orbital degrees of freedom using a slave boson representation. 
This is achieved by expanding the (pseudo)spin operators around the 
assumed mean field ground state by means of a Holstein-Primakoff 
transformation,
\begin{align}
  \label{eq:fab}
  \begin{aligned}
  d_{i \uparrow 0}^{\dag}&=f_{i}^{\dag},\\
  d_{i \downarrow 0}^{\dag}&=f_{i}^{\dag}b_{i}^{},
\end{aligned}&&
\begin{aligned}
  d_{i \uparrow 1}^{\dag}&=f_{i}^{\dag}a_{i}^{},\\
  d_{i \downarrow 1}^{\dag}&=f_{i}^{\dag}a_{i}^{}b_{i}^{},
\end{aligned}
\end{align}
where $b_i^{\dag}$ creates a~spin excitation at site $i$, $a_i^{\dag}$
creates an orbital excitation, and $f_{i}^{\dag}$ creates a spinless
fermion which represents the charge degree of freedom, where
$0$~indicates the site $i$ in the ground state, while $1$ means that
the respective site hosts an excited state. Thus, a charge can be added
to the system only if it is locally in its (classical) ground state,
otherwise if a boson occupied the considered site, first it has to be
removed before the charge can be added. Also note that the on-site
bosonic Hilbert space is restricted to $(2S+1)$ states, and since both
the spin and the pseudospin have length $\nicefrac{1}{2}$, each site 
can host not more than one boson of each kind.
This local constraint applies to every site and is fully taken into
account in our calculations through the variational technique employed.

The exchange Hamiltonian also has to be transformed into its bosonic
representation, which is done by means of the Holstein-Primakoff
transformation of the spin operators,
\begin{equation}
  \label{eq:HP}
  S_{i}^{z} = \frac{1}{2}-b_{i}^{\dag}b_{i}^{},\!\quad
  S_{i}^{+} = \sqrt{1-b_{i}^{\dag}b_{i}^{}}\,b_{i}^{},\!\quad
  S_{i}^{-} = b_{i}^{\dag} \sqrt{1-b_{i}^{\dag}b_{i}^{}},
\end{equation}
and similarly for the pseudospin operators, but in terms of orbiton
operators $\left\{a_i^{},a_i^{\dag}\right\}$. There are two issues that
are still worth pointing out concerning this transformation. Firstly,
the $S_i^z$ operators are the ones that most readily introduce higher
order terms into the Hamiltonian, and thus are principally responsible
for inter-bosonic interactions, which can have profound effects for
low-dimensional physics~\cite{Bie19} but which nonetheless are all too
often neglected in techniques re{eta}lying on the LSW
approximation. It is therefore worth mentioning that in our
variational method this part of the transformation is not strictly
necessary, as the $S_i^z$ operators merely ``count'' the Ising energy
of the bosons and thus their effect can be discerned directly from the
configuration of the system. Or, to put it differently, in our method
it would actually be more cumbersome (although possible) to calculate
the energy in the LSW approximation than it is to do it exactly.

Secondly, the square root factors in the fluctuation operators
$S_i^{\pm}$ are conventionally treated by Taylor expansion and
truncation at second order, to be consistent with the LSW approximation. 
However, here again, one should realize that these factors merely serve 
to impose the restriction of a single boson per site, and thus this 
constraint can be taken into account by excluding from the variational 
space the configurations which violate it. Therefore, with our 
variational technique we can bosonize the exchange interactions without 
the need to abandon any of the inter-bosonic interactions or constraints.

Applying these transformations decouples the original fermions into
their constituent charge, spin, and orbital degrees of freedom. The
free charge propagation $\mathcal{H}_t$, as well as its coupling to 
the bosonic degrees of freedom, will now be described by the kinetic 
Hamiltonian, which after the above transformations reads,
\begin{equation}
\mathcal{H}_t=\mathcal{T}+\mathcal{V}^{\perp}_{t}+\mathcal{V}^{\parallel}_{t}\,,
\end{equation}
where:
\begin{widetext}
  \begin{subequations}
    \label{eq:Htso4}
  \begin{align}
      \label{eq:T}
    \mathcal{T} &= -\frac{t}{4}\sum_{\mean{ij} \perp c}\left(1-2\sin\phi\right)
\left(f_{i}^{\dag}f_{j}^{} + \mathrm{H.c.}\right) =\sum_{\vect{k}}
\epsilon_{\vect{k}\phi}^{}f_{\vect{k}}^{\dag} f_{\vect{k}}^{},\\
%  \label{eq:Htso4}
    \label{eq:V1}
    \mathcal{V}^{\perp}_{t} &= \frac{t}{4} \sum_{\mean{ij} \perp c}\left\{
    \left[
2 e^{i\vect{Q}\vect{R}_{i}}\cos\phi (a_{j}^{\dag} -a_{i}^{})
   \pm\sqrt{3}(a_{j}^{\dag} +a_{i}^{})
+ (1+2\sin\phi) a_{j}^{\dag} a_{i}^{} \right]
   (1+b_{i}^{}b_{j}^{\dag}) f_{i}^{\dag} f_{j}^{} +\mathrm{H.c.}\right\}\nonumber\\
   &-\frac{t}{4}\sum_{\mean{ij}\perp c}
   \left[ (1-2\sin\phi)b_{i}^{}b_{j}^{\dag} f_{i}^{\dag} f_{j}^{}
    + \mathrm{H.c.}\right],\\
     \label{eq:V2}
    \mathcal{V}^{\parallel}_{t} &=
    -\frac{t}{2} \sum_{\mean{ij}\parallel c}\left\{
    \left[(1+\sin\phi)-e^{i\vect{Q}\vect{R}_{i}}\cos\phi 
    (a_{i}^{}+ a_{j}^{\dag})-(1-\sin\phi)a_{i}^{}a_{j}^{\dag}\right]\, 
    (b_{i}^{}+b_{j}^{\dag})f_{i}^{\dag}f_{j}^{}
    +\mathrm{H.c.}\right\},
\end{align}
\end{subequations}
\end{widetext}
and $\epsilon_{\vect{k}\phi}=-\frac12 t(1-2\sin\phi)(\cos k_x+\cos k_y)$ 
is the free electron dispersion. We emphasize that it depends on the 
orbital order (\ref{phi}) through the angle $\phi$, and vanishes when 
$\phi=\pi/6$. The free charge hopping term $\mathcal{T}$ is restricted 
to the $ab$ planes because of the magnetic order alternating in the 
perpendicular direction $c$, which thus requires creation or 
annihilation of a magnon when the charge hops in that direction. 

The remaining terms couple the charge to the bosonic degrees of freedom, 
and some of them are high-order many-body interactions, sometimes 
coupling the charge to multiple bosons at the same time. Treating these 
interactions with a technique relying on the LSW approximation would be 
impossible, whereas this can be done within the variational momentum 
average approach, as presented in Sec.~\ref{sec:res}.

\bibliographystyle{apsrev4-1}
%\bibliography{spinor}
%\bibliography{spinor.bib}

\begin{thebibliography}{68}%
\makeatletter
\providecommand \@ifxundefined [1]{%
 \@ifx{#1\undefined}
}%
\providecommand \@ifnum [1]{%
 \ifnum #1\expandafter \@firstoftwo
 \else \expandafter \@secondoftwo
 \fi
}%
\providecommand \@ifx [1]{%
 \ifx #1\expandafter \@firstoftwo
 \else \expandafter \@secondoftwo
 \fi
}%
\providecommand \natexlab [1]{#1}%
\providecommand \enquote  [1]{``#1''}%
\providecommand \bibnamefont  [1]{#1}%
\providecommand \bibfnamefont [1]{#1}%
\providecommand \citenamefont [1]{#1}%
\providecommand \href@noop [0]{\@secondoftwo}%
\providecommand \href [0]{\begingroup \@sanitize@url \@href}%
\providecommand \@href[1]{\@@startlink{#1}\@@href}%
\providecommand \@@href[1]{\endgroup#1\@@endlink}%
\providecommand \@sanitize@url [0]{\catcode `\\12\catcode `\$12\catcode
  `\&12\catcode `\#12\catcode `\^12\catcode `\_12\catcode `\%12\relax}%
\providecommand \@@startlink[1]{}%
\providecommand \@@endlink[0]{}%
\providecommand \url  [0]{\begingroup\@sanitize@url \@url }%
\providecommand \@url [1]{\endgroup\@href {#1}{\urlprefix }}%
\providecommand \urlprefix  [0]{URL }%
\providecommand \Eprint [0]{\href }%
\providecommand \doibase [0]{http://dx.doi.org/}%
\providecommand \selectlanguage [0]{\@gobble}%
\providecommand \bibinfo  [0]{\@secondoftwo}%
\providecommand \bibfield  [0]{\@secondoftwo}%
\providecommand \translation [1]{[#1]}%
\providecommand \BibitemOpen [0]{}%
\providecommand \bibitemStop [0]{}%
\providecommand \bibitemNoStop [0]{.\EOS\space}%
\providecommand \EOS [0]{\spacefactor3000\relax}%
\providecommand \BibitemShut  [1]{\csname bibitem#1\endcsname}%
\let\auto@bib@innerbib\@empty
%</preamble>
\bibitem [{\citenamefont {Chao}\ \emph {et~al.}(1977)\citenamefont {Chao},
  \citenamefont {Spa\l{}ek},\ and\ \citenamefont {Ole\'s}}]{Cha77}%
  \BibitemOpen
  \bibfield  {author} {\bibinfo {author} {\bibfnamefont {K.~A.}\ \bibnamefont
  {Chao}}, \bibinfo {author} {\bibfnamefont {J.}~\bibnamefont {Spa\l{}ek}}, \
  and\ \bibinfo {author} {\bibfnamefont {A.~M.}\ \bibnamefont {Ole\'s}},\
  }\href {\doibase 10.1088/0022-3719/10/10/002} {\bibfield  {journal} {\bibinfo
   {journal} {J. Phys. C}\ }\textbf {\bibinfo {volume} {10}},\ \bibinfo {pages}
  {L271} (\bibinfo {year} {1977})}\BibitemShut {NoStop}%
\bibitem [{\citenamefont {Trugman}(1988)}]{Tru88}%
  \BibitemOpen
  \bibfield  {author} {\bibinfo {author} {\bibfnamefont {S.~A.}\ \bibnamefont
  {Trugman}},\ }\href {\doibase 10.1103/PhysRevB.37.1597} {\bibfield  {journal}
  {\bibinfo  {journal} {Phys. Rev. B}\ }\textbf {\bibinfo {volume} {37}},\
  \bibinfo {pages} {1597} (\bibinfo {year} {1988})}\BibitemShut {NoStop}%
\bibitem [{\citenamefont {Liu}\ and\ \citenamefont {Manousakis}(1992)}]{Liu92}%
  \BibitemOpen
  \bibfield  {author} {\bibinfo {author} {\bibfnamefont {Z.}~\bibnamefont
  {Liu}}\ and\ \bibinfo {author} {\bibfnamefont {E.}~\bibnamefont
  {Manousakis}},\ }\href {\doibase 10.1103/PhysRevB.45.2425} {\bibfield
  {journal} {\bibinfo  {journal} {Phys. Rev. B}\ }\textbf {\bibinfo {volume}
  {45}},\ \bibinfo {pages} {2425} (\bibinfo {year} {1992})}\BibitemShut
  {NoStop}%
\bibitem [{\citenamefont {Manousakis}(2007)}]{Man07}%
  \BibitemOpen
  \bibfield  {author} {\bibinfo {author} {\bibfnamefont {E.}~\bibnamefont
  {Manousakis}},\ }\href {\doibase 10.1103/PhysRevB.75.035106} {\bibfield
  {journal} {\bibinfo  {journal} {Phys. Rev. B}\ }\textbf {\bibinfo {volume}
  {75}},\ \bibinfo {pages} {035106} (\bibinfo {year} {2007})}\BibitemShut
  {NoStop}%
\bibitem [{\citenamefont {Grusdt}\ \emph {et~al.}(2018)\citenamefont {Grusdt},
  \citenamefont {K\'anasz-Nagy}, \citenamefont {Bohrdt}, \citenamefont {Chiu},
  \citenamefont {Ji}, \citenamefont {Greiner}, \citenamefont {Greif},\ and\
  \citenamefont {Demler}}]{Gru18}%
  \BibitemOpen
  \bibfield  {author} {\bibinfo {author} {\bibfnamefont {F.}~\bibnamefont
  {Grusdt}}, \bibinfo {author} {\bibfnamefont {M.}~\bibnamefont
  {K\'anasz-Nagy}}, \bibinfo {author} {\bibfnamefont {A.}~\bibnamefont
  {Bohrdt}}, \bibinfo {author} {\bibfnamefont {C.~S.}\ \bibnamefont {Chiu}},
  \bibinfo {author} {\bibfnamefont {G.}~\bibnamefont {Ji}}, \bibinfo {author}
  {\bibfnamefont {M.}~\bibnamefont {Greiner}}, \bibinfo {author} {\bibfnamefont
  {D.}~\bibnamefont {Greif}}, \ and\ \bibinfo {author} {\bibfnamefont
  {E.}~\bibnamefont {Demler}},\ }\href {\doibase 10.1103/PhysRevX.8.011046}
  {\bibfield  {journal} {\bibinfo  {journal} {Phys. Rev. X}\ }\textbf {\bibinfo
  {volume} {8}},\ \bibinfo {pages} {011046} (\bibinfo {year}
  {2018})}\BibitemShut {NoStop}%
\bibitem [{\citenamefont {Kane}\ \emph {et~al.}(1989)\citenamefont {Kane},
  \citenamefont {Lee},\ and\ \citenamefont {Read}}]{Kan89}%
  \BibitemOpen
  \bibfield  {author} {\bibinfo {author} {\bibfnamefont {C.~L.}\ \bibnamefont
  {Kane}}, \bibinfo {author} {\bibfnamefont {P.~A.}\ \bibnamefont {Lee}}, \
  and\ \bibinfo {author} {\bibfnamefont {N.}~\bibnamefont {Read}},\ }\href
  {\doibase 10.1103/PhysRevB.39.6880} {\bibfield  {journal} {\bibinfo
  {journal} {Phys. Rev. B}\ }\textbf {\bibinfo {volume} {39}},\ \bibinfo
  {pages} {6880} (\bibinfo {year} {1989})}\BibitemShut {NoStop}%
\bibitem [{\citenamefont {Mart\'inez}\ and\ \citenamefont
  {Horsch}(1991)}]{Mar91}%
  \BibitemOpen
  \bibfield  {author} {\bibinfo {author} {\bibfnamefont {G.}~\bibnamefont
  {Mart\'inez}}\ and\ \bibinfo {author} {\bibfnamefont {P.}~\bibnamefont
  {Horsch}},\ }\href {\doibase 10.1103/PhysRevB.44.317} {\bibfield  {journal}
  {\bibinfo  {journal} {Phys. Rev. B}\ }\textbf {\bibinfo {volume} {44}},\
  \bibinfo {pages} {317} (\bibinfo {year} {1991})}\BibitemShut {NoStop}%
\bibitem [{\citenamefont {Lee}\ \emph {et~al.}(2006)\citenamefont {Lee},
  \citenamefont {Nagaosa},\ and\ \citenamefont {Wen}}]{Lee06}%
  \BibitemOpen
  \bibfield  {author} {\bibinfo {author} {\bibfnamefont {P.~A.}\ \bibnamefont
  {Lee}}, \bibinfo {author} {\bibfnamefont {N.}~\bibnamefont {Nagaosa}}, \ and\
  \bibinfo {author} {\bibfnamefont {X.-G.}\ \bibnamefont {Wen}},\ }\href
  {\doibase 10.1103/RevModPhys.78.17} {\bibfield  {journal} {\bibinfo
  {journal} {Rev. Mod. Phys.}\ }\textbf {\bibinfo {volume} {78}},\ \bibinfo
  {pages} {17} (\bibinfo {year} {2006})}\BibitemShut {NoStop}%
\bibitem [{\citenamefont {Kugel}\ and\ \citenamefont {Khomskii}(1982)}]{Kug82}%
  \BibitemOpen
  \bibfield  {author} {\bibinfo {author} {\bibfnamefont {K.~I.}\ \bibnamefont
  {Kugel}}\ and\ \bibinfo {author} {\bibfnamefont {D.~I.}\ \bibnamefont
  {Khomskii}},\ }\href {\doibase 10.1070/PU1982v025n04ABEH004537} {\bibfield
  {journal} {\bibinfo  {journal} {Sov. Phys. Usp.}\ }\textbf {\bibinfo {volume}
  {25}},\ \bibinfo {pages} {231} (\bibinfo {year} {1982})}\BibitemShut
  {NoStop}%
\bibitem [{\citenamefont {Tokura}\ and\ \citenamefont {Nagaosa}(2000)}]{Tok00}%
  \BibitemOpen
  \bibfield  {author} {\bibinfo {author} {\bibfnamefont {Y.}~\bibnamefont
  {Tokura}}\ and\ \bibinfo {author} {\bibfnamefont {N.}~\bibnamefont
  {Nagaosa}},\ }\href {\doibase 10.1126/science.288.5465.462} {\bibfield
  {journal} {\bibinfo  {journal} {Science}\ }\textbf {\bibinfo {volume}
  {288}},\ \bibinfo {pages} {462} (\bibinfo {year} {2000})}\BibitemShut
  {NoStop}%
\bibitem [{\citenamefont {Ole\ifmmode~\acute{s}\else
  \'{s}\fi{}}(1983)}]{Ole83}%
  \BibitemOpen
  \bibfield  {author} {\bibinfo {author} {\bibfnamefont {A.~M.}\ \bibnamefont
  {Ole\ifmmode~\acute{s}\else \'{s}\fi{}}},\ }\href {\doibase
  10.1103/PhysRevB.28.327} {\bibfield  {journal} {\bibinfo  {journal} {Phys.
  Rev. B}\ }\textbf {\bibinfo {volume} {28}},\ \bibinfo {pages} {327} (\bibinfo
  {year} {1983})}\BibitemShut {NoStop}%
\bibitem [{\citenamefont {Hoshino}\ and\ \citenamefont {Werner}(2016)}]{Hos16}%
  \BibitemOpen
  \bibfield  {author} {\bibinfo {author} {\bibfnamefont {S.}~\bibnamefont
  {Hoshino}}\ and\ \bibinfo {author} {\bibfnamefont {P.}~\bibnamefont
  {Werner}},\ }\href {\doibase 10.1103/PhysRevB.93.155161} {\bibfield
  {journal} {\bibinfo  {journal} {Phys. Rev. B}\ }\textbf {\bibinfo {volume}
  {93}},\ \bibinfo {pages} {155161} (\bibinfo {year} {2016})}\BibitemShut
  {NoStop}%
\bibitem [{\citenamefont {Feiner}\ \emph {et~al.}(1997)\citenamefont {Feiner},
  \citenamefont {Ole\ifmmode~\acute{s}\else \'{s}\fi{}},\ and\ \citenamefont
  {Zaanen}}]{Fei97}%
  \BibitemOpen
  \bibfield  {author} {\bibinfo {author} {\bibfnamefont {L.~F.}\ \bibnamefont
  {Feiner}}, \bibinfo {author} {\bibfnamefont {A.~M.}\ \bibnamefont
  {Ole\ifmmode~\acute{s}\else \'{s}\fi{}}}, \ and\ \bibinfo {author}
  {\bibfnamefont {J.}~\bibnamefont {Zaanen}},\ }\href {\doibase
  10.1103/PhysRevLett.78.2799} {\bibfield  {journal} {\bibinfo  {journal}
  {Phys. Rev. Lett.}\ }\textbf {\bibinfo {volume} {78}},\ \bibinfo {pages}
  {2799} (\bibinfo {year} {1997})}\BibitemShut {NoStop}%
\bibitem [{\citenamefont {Ishihara}\ \emph {et~al.}(1997)\citenamefont
  {Ishihara}, \citenamefont {Inoue},\ and\ \citenamefont {Maekawa}}]{Ish97}%
  \BibitemOpen
  \bibfield  {author} {\bibinfo {author} {\bibfnamefont {S.}~\bibnamefont
  {Ishihara}}, \bibinfo {author} {\bibfnamefont {J.}~\bibnamefont {Inoue}}, \
  and\ \bibinfo {author} {\bibfnamefont {S.}~\bibnamefont {Maekawa}},\ }\href
  {\doibase 10.1103/PhysRevB.55.8280} {\bibfield  {journal} {\bibinfo
  {journal} {Phys. Rev. B}\ }\textbf {\bibinfo {volume} {55}},\ \bibinfo
  {pages} {8280} (\bibinfo {year} {1997})}\BibitemShut {NoStop}%
\bibitem [{\citenamefont {Feiner}\ and\ \citenamefont
  {Ole\ifmmode~\acute{s}\else \'{s}\fi{}}(1999)}]{Fei99}%
  \BibitemOpen
  \bibfield  {author} {\bibinfo {author} {\bibfnamefont {L.~F.}\ \bibnamefont
  {Feiner}}\ and\ \bibinfo {author} {\bibfnamefont {A.~M.}\ \bibnamefont
  {Ole\ifmmode~\acute{s}\else \'{s}\fi{}}},\ }\href {\doibase
  10.1103/PhysRevB.59.3295} {\bibfield  {journal} {\bibinfo  {journal} {Phys.
  Rev. B}\ }\textbf {\bibinfo {volume} {59}},\ \bibinfo {pages} {3295}
  (\bibinfo {year} {1999})}\BibitemShut {NoStop}%
\bibitem [{\citenamefont {Snamina}\ and\ \citenamefont
  {Ole\ifmmode~\acute{s}\else \'{s}\fi{}}(2018)}]{Sna18}%
  \BibitemOpen
  \bibfield  {author} {\bibinfo {author} {\bibfnamefont {M.}~\bibnamefont
  {Snamina}}\ and\ \bibinfo {author} {\bibfnamefont {A.~M.}\ \bibnamefont
  {Ole\ifmmode~\acute{s}\else \'{s}\fi{}}},\ }\href {\doibase
  10.1103/PhysRevB.97.104417} {\bibfield  {journal} {\bibinfo  {journal} {Phys.
  Rev. B}\ }\textbf {\bibinfo {volume} {97}},\ \bibinfo {pages} {104417}
  (\bibinfo {year} {2018})}\BibitemShut {NoStop}%
\bibitem [{\citenamefont {Khaliullin}\ and\ \citenamefont
  {Maekawa}(2000)}]{Kha00}%
  \BibitemOpen
  \bibfield  {author} {\bibinfo {author} {\bibfnamefont {G.}~\bibnamefont
  {Khaliullin}}\ and\ \bibinfo {author} {\bibfnamefont {S.}~\bibnamefont
  {Maekawa}},\ }\href {\doibase 10.1103/PhysRevLett.85.3950} {\bibfield
  {journal} {\bibinfo  {journal} {Phys. Rev. Lett.}\ }\textbf {\bibinfo
  {volume} {85}},\ \bibinfo {pages} {3950} (\bibinfo {year}
  {2000})}\BibitemShut {NoStop}%
\bibitem [{\citenamefont {Khaliullin}\ \emph {et~al.}(2001)\citenamefont
  {Khaliullin}, \citenamefont {Horsch},\ and\ \citenamefont
  {Ole\ifmmode~\acute{s}\else \'{s}\fi{}}}]{Kha01}%
  \BibitemOpen
  \bibfield  {author} {\bibinfo {author} {\bibfnamefont {G.}~\bibnamefont
  {Khaliullin}}, \bibinfo {author} {\bibfnamefont {P.}~\bibnamefont {Horsch}},
  \ and\ \bibinfo {author} {\bibfnamefont {A.~M.}\ \bibnamefont
  {Ole\ifmmode~\acute{s}\else \'{s}\fi{}}},\ }\href {\doibase
  10.1103/PhysRevLett.86.3879} {\bibfield  {journal} {\bibinfo  {journal}
  {Phys. Rev. Lett.}\ }\textbf {\bibinfo {volume} {86}},\ \bibinfo {pages}
  {3879} (\bibinfo {year} {2001})}\BibitemShut {NoStop}%
\bibitem [{\citenamefont {Khaliullin}\ \emph {et~al.}(2004)\citenamefont
  {Khaliullin}, \citenamefont {Horsch},\ and\ \citenamefont
  {Ole\ifmmode~\acute{s}\else \'{s}\fi{}}}]{Kha04}%
  \BibitemOpen
  \bibfield  {author} {\bibinfo {author} {\bibfnamefont {G.}~\bibnamefont
  {Khaliullin}}, \bibinfo {author} {\bibfnamefont {P.}~\bibnamefont {Horsch}},
  \ and\ \bibinfo {author} {\bibfnamefont {A.~M.}\ \bibnamefont
  {Ole\ifmmode~\acute{s}\else \'{s}\fi{}}},\ }\href {\doibase
  10.1103/PhysRevB.70.195103} {\bibfield  {journal} {\bibinfo  {journal} {Phys.
  Rev. B}\ }\textbf {\bibinfo {volume} {70}},\ \bibinfo {pages} {195103}
  (\bibinfo {year} {2004})}\BibitemShut {NoStop}%
\bibitem [{\citenamefont {Khaliullin}(2005)}]{Kha05}%
  \BibitemOpen
  \bibfield  {author} {\bibinfo {author} {\bibfnamefont {G.}~\bibnamefont
  {Khaliullin}},\ }\href {\doibase 10.1143/PTPS.160.155} {\bibfield  {journal}
  {\bibinfo  {journal} {Prog. Theor. Phys. Suppl.}\ }\textbf {\bibinfo {volume}
  {160}},\ \bibinfo {pages} {155} (\bibinfo {year} {2005})}\BibitemShut
  {NoStop}%
\bibitem [{\citenamefont {Ole\'s}\ \emph {et~al.}(2005)\citenamefont {Ole\'s},
  \citenamefont {Khaliullin}, \citenamefont {Horsch},\ and\ \citenamefont
  {Feiner}}]{Ole05}%
  \BibitemOpen
  \bibfield  {author} {\bibinfo {author} {\bibfnamefont {A.~M.}\ \bibnamefont
  {Ole\'s}}, \bibinfo {author} {\bibfnamefont {G.}~\bibnamefont {Khaliullin}},
  \bibinfo {author} {\bibfnamefont {P.}~\bibnamefont {Horsch}}, \ and\ \bibinfo
  {author} {\bibfnamefont {L.~F.}\ \bibnamefont {Feiner}},\ }\href {\doibase
  10.1103/PhysRevB.72.214431} {\bibfield  {journal} {\bibinfo  {journal} {Phys.
  Rev. B}\ }\textbf {\bibinfo {volume} {72}},\ \bibinfo {pages} {214431}
  (\bibinfo {year} {2005})}\BibitemShut {NoStop}%
\bibitem [{\citenamefont {Chaloupka}\ and\ \citenamefont
  {Khaliullin}(2008)}]{Cha08}%
  \BibitemOpen
  \bibfield  {author} {\bibinfo {author} {\bibfnamefont {J.}~\bibnamefont
  {Chaloupka}}\ and\ \bibinfo {author} {\bibfnamefont {G.}~\bibnamefont
  {Khaliullin}},\ }\href {\doibase 10.1103/PhysRevLett.100.016404} {\bibfield
  {journal} {\bibinfo  {journal} {Phys. Rev. Lett.}\ }\textbf {\bibinfo
  {volume} {100}},\ \bibinfo {pages} {016404} (\bibinfo {year}
  {2008})}\BibitemShut {NoStop}%
\bibitem [{\citenamefont {Horsch}\ \emph {et~al.}(2008)\citenamefont {Horsch},
  \citenamefont {Ole\ifmmode~\acute{s}\else \'{s}\fi{}}, \citenamefont
  {Feiner},\ and\ \citenamefont {Khaliullin}}]{Hor08}%
  \BibitemOpen
  \bibfield  {author} {\bibinfo {author} {\bibfnamefont {P.}~\bibnamefont
  {Horsch}}, \bibinfo {author} {\bibfnamefont {A.~M.}\ \bibnamefont
  {Ole\ifmmode~\acute{s}\else \'{s}\fi{}}}, \bibinfo {author} {\bibfnamefont
  {L.~F.}\ \bibnamefont {Feiner}}, \ and\ \bibinfo {author} {\bibfnamefont
  {G.}~\bibnamefont {Khaliullin}},\ }\href {\doibase
  10.1103/PhysRevLett.100.167205} {\bibfield  {journal} {\bibinfo  {journal}
  {Phys. Rev. Lett.}\ }\textbf {\bibinfo {volume} {100}},\ \bibinfo {pages}
  {167205} (\bibinfo {year} {2008})}\BibitemShut {NoStop}%
\bibitem [{\citenamefont {Normand}\ and\ \citenamefont
  {Ole\ifmmode~\acute{s}\else \'{s}\fi{}}(2008)}]{Nor08}%
  \BibitemOpen
  \bibfield  {author} {\bibinfo {author} {\bibfnamefont {B.}~\bibnamefont
  {Normand}}\ and\ \bibinfo {author} {\bibfnamefont {A.~M.}\ \bibnamefont
  {Ole\ifmmode~\acute{s}\else \'{s}\fi{}}},\ }\href {\doibase
  10.1103/PhysRevB.78.094427} {\bibfield  {journal} {\bibinfo  {journal} {Phys.
  Rev. B}\ }\textbf {\bibinfo {volume} {78}},\ \bibinfo {pages} {094427}
  (\bibinfo {year} {2008})}\BibitemShut {NoStop}%
\bibitem [{\citenamefont {Normand}(2011)}]{Nor11}%
  \BibitemOpen
  \bibfield  {author} {\bibinfo {author} {\bibfnamefont {B.}~\bibnamefont
  {Normand}},\ }\href {\doibase 10.1103/PhysRevB.83.064413} {\bibfield
  {journal} {\bibinfo  {journal} {Phys. Rev. B}\ }\textbf {\bibinfo {volume}
  {83}},\ \bibinfo {pages} {064413} (\bibinfo {year} {2011})}\BibitemShut
  {NoStop}%
\bibitem [{\citenamefont {Chaloupka}\ and\ \citenamefont
  {Ole\ifmmode~\acute{s}\else \'{s}\fi{}}(2011)}]{Cha11}%
  \BibitemOpen
  \bibfield  {author} {\bibinfo {author} {\bibfnamefont {J.}~\bibnamefont
  {Chaloupka}}\ and\ \bibinfo {author} {\bibfnamefont {A.~M.}\ \bibnamefont
  {Ole\ifmmode~\acute{s}\else \'{s}\fi{}}},\ }\href {\doibase
  10.1103/PhysRevB.83.094406} {\bibfield  {journal} {\bibinfo  {journal} {Phys.
  Rev. B}\ }\textbf {\bibinfo {volume} {83}},\ \bibinfo {pages} {094406}
  (\bibinfo {year} {2011})}\BibitemShut {NoStop}%
\bibitem [{\citenamefont {Sirker}\ \emph {et~al.}(2008)\citenamefont {Sirker},
  \citenamefont {Herzog}, \citenamefont {Ole\ifmmode~\acute{s}\else
  \'{s}\fi{}},\ and\ \citenamefont {Horsch}}]{Sir08}%
  \BibitemOpen
  \bibfield  {author} {\bibinfo {author} {\bibfnamefont {J.}~\bibnamefont
  {Sirker}}, \bibinfo {author} {\bibfnamefont {A.}~\bibnamefont {Herzog}},
  \bibinfo {author} {\bibfnamefont {A.~M.}\ \bibnamefont
  {Ole\ifmmode~\acute{s}\else \'{s}\fi{}}}, \ and\ \bibinfo {author}
  {\bibfnamefont {P.}~\bibnamefont {Horsch}},\ }\href {\doibase
  10.1103/PhysRevLett.101.157204} {\bibfield  {journal} {\bibinfo  {journal}
  {Phys. Rev. Lett.}\ }\textbf {\bibinfo {volume} {101}},\ \bibinfo {pages}
  {157204} (\bibinfo {year} {2008})}\BibitemShut {NoStop}%
\bibitem [{\citenamefont {Herzog}\ \emph {et~al.}(2011)\citenamefont {Herzog},
  \citenamefont {Horsch}, \citenamefont {Ole\ifmmode~\acute{s}\else
  \'{s}\fi{}},\ and\ \citenamefont {Sirker}}]{Her11}%
  \BibitemOpen
  \bibfield  {author} {\bibinfo {author} {\bibfnamefont {A.}~\bibnamefont
  {Herzog}}, \bibinfo {author} {\bibfnamefont {P.}~\bibnamefont {Horsch}},
  \bibinfo {author} {\bibfnamefont {A.~M.}\ \bibnamefont
  {Ole\ifmmode~\acute{s}\else \'{s}\fi{}}}, \ and\ \bibinfo {author}
  {\bibfnamefont {J.}~\bibnamefont {Sirker}},\ }\href {\doibase
  10.1103/PhysRevB.83.245130} {\bibfield  {journal} {\bibinfo  {journal} {Phys.
  Rev. B}\ }\textbf {\bibinfo {volume} {83}},\ \bibinfo {pages} {245130}
  (\bibinfo {year} {2011})}\BibitemShut {NoStop}%
\bibitem [{\citenamefont {Brzezicki}\ \emph {et~al.}(2015)\citenamefont
  {Brzezicki}, \citenamefont {Ole\ifmmode~\acute{s}\else \'{s}\fi{}},\ and\
  \citenamefont {Cuoco}}]{Brz15}%
  \BibitemOpen
  \bibfield  {author} {\bibinfo {author} {\bibfnamefont {W.}~\bibnamefont
  {Brzezicki}}, \bibinfo {author} {\bibfnamefont {A.~M.}\ \bibnamefont
  {Ole\ifmmode~\acute{s}\else \'{s}\fi{}}}, \ and\ \bibinfo {author}
  {\bibfnamefont {M.}~\bibnamefont {Cuoco}},\ }\href {\doibase
  10.1103/PhysRevX.5.011037} {\bibfield  {journal} {\bibinfo  {journal} {Phys.
  Rev. X}\ }\textbf {\bibinfo {volume} {5}},\ \bibinfo {pages} {011037}
  (\bibinfo {year} {2015})}\BibitemShut {NoStop}%
\bibitem [{\citenamefont {Brzezicki}(2019)}]{Brz19}%
  \BibitemOpen
  \bibfield  {author} {\bibinfo {author} {\bibfnamefont {W.}~\bibnamefont
  {Brzezicki}},\ }\href@noop {} {\bibfield  {journal} {\bibinfo  {journal}
  {arXiv:1904.11772}\ } (\bibinfo {year} {2019})}\BibitemShut {NoStop}%
\bibitem [{\citenamefont {van~den Brink}\ \emph {et~al.}(2000)\citenamefont
  {van~den Brink}, \citenamefont {Horsch},\ and\ \citenamefont
  {Ole\'s}}]{vdB00}%
  \BibitemOpen
  \bibfield  {author} {\bibinfo {author} {\bibfnamefont {J.}~\bibnamefont
  {van~den Brink}}, \bibinfo {author} {\bibfnamefont {P.}~\bibnamefont
  {Horsch}}, \ and\ \bibinfo {author} {\bibfnamefont {A.~M.}\ \bibnamefont
  {Ole\'s}},\ }\href {\doibase 10.1103/PhysRevLett.85.5174} {\bibfield
  {journal} {\bibinfo  {journal} {Phys. Rev. Lett.}\ }\textbf {\bibinfo
  {volume} {85}},\ \bibinfo {pages} {5174} (\bibinfo {year}
  {2000})}\BibitemShut {NoStop}%
\bibitem [{\citenamefont {Ishihara}\ \emph {et~al.}(2005)\citenamefont
  {Ishihara}, \citenamefont {Murakami}, \citenamefont {Inami}, \citenamefont
  {Ishii}, \citenamefont {Mizuki}, \citenamefont {Hirota}, \citenamefont
  {Maekawa},\ and\ \citenamefont {Endoh}}]{Ish05}%
  \BibitemOpen
  \bibfield  {author} {\bibinfo {author} {\bibfnamefont {S.}~\bibnamefont
  {Ishihara}}, \bibinfo {author} {\bibfnamefont {Y.}~\bibnamefont {Murakami}},
  \bibinfo {author} {\bibfnamefont {T.}~\bibnamefont {Inami}}, \bibinfo
  {author} {\bibfnamefont {K.}~\bibnamefont {Ishii}}, \bibinfo {author}
  {\bibfnamefont {J.}~\bibnamefont {Mizuki}}, \bibinfo {author} {\bibfnamefont
  {K.}~\bibnamefont {Hirota}}, \bibinfo {author} {\bibfnamefont
  {S.}~\bibnamefont {Maekawa}}, \ and\ \bibinfo {author} {\bibfnamefont
  {Y.}~\bibnamefont {Endoh}},\ }\href
  {http://stacks.iop.org/1367-2630/7/i=1/a=119} {\bibfield  {journal} {\bibinfo
   {journal} {New J. Phys.}\ }\textbf {\bibinfo {volume} {7}},\ \bibinfo
  {pages} {119} (\bibinfo {year} {2005})}\BibitemShut {NoStop}%
\bibitem [{\citenamefont {Ishihara}(2005)}]{Ishih}%
  \BibitemOpen
  \bibfield  {author} {\bibinfo {author} {\bibfnamefont {S.}~\bibnamefont
  {Ishihara}},\ }\href {\doibase 10.1103/PhysRevLett.94.156408} {\bibfield
  {journal} {\bibinfo  {journal} {Phys. Rev. Lett.}\ }\textbf {\bibinfo
  {volume} {94}},\ \bibinfo {pages} {156408} (\bibinfo {year}
  {2005})}\BibitemShut {NoStop}%
\bibitem [{\citenamefont {Daghofer}\ \emph {et~al.}(2004)\citenamefont
  {Daghofer}, \citenamefont {Ole\'s},\ and\ \citenamefont {von~der
  Linden}}]{Dag04}%
  \BibitemOpen
  \bibfield  {author} {\bibinfo {author} {\bibfnamefont {M.}~\bibnamefont
  {Daghofer}}, \bibinfo {author} {\bibfnamefont {A.~M.}\ \bibnamefont
  {Ole\'s}}, \ and\ \bibinfo {author} {\bibfnamefont {W.}~\bibnamefont {von~der
  Linden}},\ }\href {\doibase 10.1103/PhysRevB.70.184430} {\bibfield  {journal}
  {\bibinfo  {journal} {Phys. Rev. B}\ }\textbf {\bibinfo {volume} {70}},\
  \bibinfo {pages} {184430} (\bibinfo {year} {2004})}\BibitemShut {NoStop}%
\bibitem [{\citenamefont {Wohlfeld}\ \emph {et~al.}(2009)\citenamefont
  {Wohlfeld}, \citenamefont {Ole\ifmmode~\acute{s}\else \'{s}\fi{}},\ and\
  \citenamefont {Horsch}}]{Woh09}%
  \BibitemOpen
  \bibfield  {author} {\bibinfo {author} {\bibfnamefont {K.}~\bibnamefont
  {Wohlfeld}}, \bibinfo {author} {\bibfnamefont {A.~M.}\ \bibnamefont
  {Ole\ifmmode~\acute{s}\else \'{s}\fi{}}}, \ and\ \bibinfo {author}
  {\bibfnamefont {P.}~\bibnamefont {Horsch}},\ }\href {\doibase
  10.1103/PhysRevB.79.224433} {\bibfield  {journal} {\bibinfo  {journal} {Phys.
  Rev. B}\ }\textbf {\bibinfo {volume} {79}},\ \bibinfo {pages} {224433}
  (\bibinfo {year} {2009})}\BibitemShut {NoStop}%
\bibitem [{\citenamefont {Berciu}(2009)}]{Ber09a}%
  \BibitemOpen
  \bibfield  {author} {\bibinfo {author} {\bibfnamefont {M.}~\bibnamefont
  {Berciu}},\ }\href {\doibase 10.1103/Physics.2.55} {\bibfield  {journal}
  {\bibinfo  {journal} {Physics}\ }\textbf {\bibinfo {volume} {2}},\ \bibinfo
  {pages} {55} (\bibinfo {year} {2009})}\BibitemShut {NoStop}%
\bibitem [{\citenamefont {Okazaki}\ and\ \citenamefont
  {Suemune}(1961)}]{Oka61}%
  \BibitemOpen
  \bibfield  {author} {\bibinfo {author} {\bibfnamefont {A.}~\bibnamefont
  {Okazaki}}\ and\ \bibinfo {author} {\bibfnamefont {Y.}~\bibnamefont
  {Suemune}},\ }\href {\doibase 10.1143/JPSJ.16.176} {\bibfield  {journal}
  {\bibinfo  {journal} {J. Phys. Soc. Japan}\ }\textbf {\bibinfo {volume}
  {16}},\ \bibinfo {pages} {176} (\bibinfo {year} {1961})}\BibitemShut
  {NoStop}%
\bibitem [{\citenamefont {Zhou}\ and\ \citenamefont
  {Goodenough}(2006)}]{Zho06}%
  \BibitemOpen
  \bibfield  {author} {\bibinfo {author} {\bibfnamefont {J.-S.}\ \bibnamefont
  {Zhou}}\ and\ \bibinfo {author} {\bibfnamefont {J.~B.}\ \bibnamefont
  {Goodenough}},\ }\href {\doibase 10.1103/PhysRevLett.96.247202} {\bibfield
  {journal} {\bibinfo  {journal} {Phys. Rev. Lett.}\ }\textbf {\bibinfo
  {volume} {96}},\ \bibinfo {pages} {247202} (\bibinfo {year}
  {2006})}\BibitemShut {NoStop}%
\bibitem [{\citenamefont {Kimura}\ \emph {et~al.}(2003)\citenamefont {Kimura},
  \citenamefont {Ishihara}, \citenamefont {Shintani}, \citenamefont {Arima},
  \citenamefont {Takahashi}, \citenamefont {Ishizaka},\ and\ \citenamefont
  {Tokura}}]{Kim03}%
  \BibitemOpen
  \bibfield  {author} {\bibinfo {author} {\bibfnamefont {T.}~\bibnamefont
  {Kimura}}, \bibinfo {author} {\bibfnamefont {S.}~\bibnamefont {Ishihara}},
  \bibinfo {author} {\bibfnamefont {H.}~\bibnamefont {Shintani}}, \bibinfo
  {author} {\bibfnamefont {T.}~\bibnamefont {Arima}}, \bibinfo {author}
  {\bibfnamefont {K.~T.}\ \bibnamefont {Takahashi}}, \bibinfo {author}
  {\bibfnamefont {K.}~\bibnamefont {Ishizaka}}, \ and\ \bibinfo {author}
  {\bibfnamefont {Y.}~\bibnamefont {Tokura}},\ }\href {\doibase
  10.1103/PhysRevB.68.060403} {\bibfield  {journal} {\bibinfo  {journal} {Phys.
  Rev. B}\ }\textbf {\bibinfo {volume} {68}},\ \bibinfo {pages} {060403}
  (\bibinfo {year} {2003})}\BibitemShut {NoStop}%
\bibitem [{\citenamefont {Lake}\ \emph
  {et~al.}(2005{\natexlab{a}})\citenamefont {Lake}, \citenamefont {Tennant},
  \citenamefont {Frost},\ and\ \citenamefont {Nagler}}]{Lak05}%
  \BibitemOpen
  \bibfield  {author} {\bibinfo {author} {\bibfnamefont {B.}~\bibnamefont
  {Lake}}, \bibinfo {author} {\bibfnamefont {D.~A.}\ \bibnamefont {Tennant}},
  \bibinfo {author} {\bibfnamefont {C.~D.}\ \bibnamefont {Frost}}, \ and\
  \bibinfo {author} {\bibfnamefont {S.~E.}\ \bibnamefont {Nagler}},\ }\href
  {\doibase 10.1038/nmat1327} {\bibfield  {journal} {\bibinfo  {journal}
  {Nature Mat.}\ }\textbf {\bibinfo {volume} {4}},\ \bibinfo {pages} {329}
  (\bibinfo {year} {2005}{\natexlab{a}})}\BibitemShut {NoStop}%
\bibitem [{\citenamefont {Lake}\ \emph
  {et~al.}(2005{\natexlab{b}})\citenamefont {Lake}, \citenamefont {Tennant},\
  and\ \citenamefont {Nagler}}]{Lak05a}%
  \BibitemOpen
  \bibfield  {author} {\bibinfo {author} {\bibfnamefont {B.}~\bibnamefont
  {Lake}}, \bibinfo {author} {\bibfnamefont {D.~A.}\ \bibnamefont {Tennant}}, \
  and\ \bibinfo {author} {\bibfnamefont {S.~E.}\ \bibnamefont {Nagler}},\
  }\href {\doibase 10.1103/PhysRevB.71.134412} {\bibfield  {journal} {\bibinfo
  {journal} {Phys. Rev. B}\ }\textbf {\bibinfo {volume} {71}},\ \bibinfo
  {pages} {134412} (\bibinfo {year} {2005}{\natexlab{b}})}\BibitemShut
  {NoStop}%
\bibitem [{\citenamefont {{Jonker}}\ and\ \citenamefont {{Van
  Santen}}(1950)}]{Jon50}%
  \BibitemOpen
  \bibfield  {author} {\bibinfo {author} {\bibfnamefont {G.~H.}\ \bibnamefont
  {{Jonker}}}\ and\ \bibinfo {author} {\bibfnamefont {J.~H.}\ \bibnamefont
  {{van Santen}}},\ }\href {\doibase 10.1016/0031-8914(50)90033-4} {\bibfield
  {journal} {\bibinfo  {journal} {Physica}\ }\textbf {\bibinfo {volume} {16}},\
  \bibinfo {pages} {337} (\bibinfo {year} {1950})}\BibitemShut {NoStop}%
\bibitem [{\citenamefont {Tokura}(2006)}]{Tok06}%
  \BibitemOpen
  \bibfield  {author} {\bibinfo {author} {\bibfnamefont {Y.}~\bibnamefont
  {Tokura}},\ }\href {\doibase 10.1088/0034-4885/69/3/r06} {\bibfield
  {journal} {\bibinfo  {journal} {Reports on Progress in Physics}\ }\textbf
  {\bibinfo {volume} {69}},\ \bibinfo {pages} {797} (\bibinfo {year}
  {2006})}\BibitemShut {NoStop}%
\bibitem [{\citenamefont {Ro\ifmmode~\acute{s}\else \'{s}\fi{}ciszewski}\ and\
  \citenamefont {Ole\ifmmode~\acute{s}\else \'{s}\fi{}}(2019)}]{Ros19}%
  \BibitemOpen
  \bibfield  {author} {\bibinfo {author} {\bibfnamefont {K.}~\bibnamefont
  {Ro\ifmmode~\acute{s}\else \'{s}\fi{}ciszewski}}\ and\ \bibinfo {author}
  {\bibfnamefont {A.~M.}\ \bibnamefont {Ole\ifmmode~\acute{s}\else
  \'{s}\fi{}}},\ }\href {\doibase 10.1103/PhysRevB.99.155108} {\bibfield
  {journal} {\bibinfo  {journal} {Phys. Rev. B}\ }\textbf {\bibinfo {volume}
  {99}},\ \bibinfo {pages} {155108} (\bibinfo {year} {2019})}\BibitemShut
  {NoStop}%
\bibitem [{\citenamefont {Ba\l{}a}\ \emph {et~al.}(2001)\citenamefont
  {Ba\l{}a}, \citenamefont {Sawatzky}, \citenamefont
  {Ole\ifmmode~\acute{s}\else \'{s}\fi{}},\ and\ \citenamefont
  {Macridin}}]{Bal01}%
  \BibitemOpen
  \bibfield  {author} {\bibinfo {author} {\bibfnamefont {J.}~\bibnamefont
  {Ba\l{}a}}, \bibinfo {author} {\bibfnamefont {G.~A.}\ \bibnamefont
  {Sawatzky}}, \bibinfo {author} {\bibfnamefont {A.~M.}\ \bibnamefont
  {Ole\ifmmode~\acute{s}\else \'{s}\fi{}}}, \ and\ \bibinfo {author}
  {\bibfnamefont {A.}~\bibnamefont {Macridin}},\ }\href {\doibase
  10.1103/PhysRevLett.87.067204} {\bibfield  {journal} {\bibinfo  {journal}
  {Phys. Rev. Lett.}\ }\textbf {\bibinfo {volume} {87}},\ \bibinfo {pages}
  {067204} (\bibinfo {year} {2001})}\BibitemShut {NoStop}%
\bibitem [{\citenamefont {Kr\"uger}\ \emph {et~al.}(2004)\citenamefont
  {Kr\"uger}, \citenamefont {Schulz}, \citenamefont {Naler}, \citenamefont
  {Rauer}, \citenamefont {Budelmann}, \citenamefont {B\"ackstr\"om},
  \citenamefont {Kim}, \citenamefont {Cheong}, \citenamefont {Perebeinos},\
  and\ \citenamefont {R\"ubhausen}}]{Kru04}%
  \BibitemOpen
  \bibfield  {author} {\bibinfo {author} {\bibfnamefont {R.}~\bibnamefont
  {Kr\"uger}}, \bibinfo {author} {\bibfnamefont {B.}~\bibnamefont {Schulz}},
  \bibinfo {author} {\bibfnamefont {S.}~\bibnamefont {Naler}}, \bibinfo
  {author} {\bibfnamefont {R.}~\bibnamefont {Rauer}}, \bibinfo {author}
  {\bibfnamefont {D.}~\bibnamefont {Budelmann}}, \bibinfo {author}
  {\bibfnamefont {J.}~\bibnamefont {B\"ackstr\"om}}, \bibinfo {author}
  {\bibfnamefont {K.~H.}\ \bibnamefont {Kim}}, \bibinfo {author} {\bibfnamefont
  {S.-W.}\ \bibnamefont {Cheong}}, \bibinfo {author} {\bibfnamefont
  {V.}~\bibnamefont {Perebeinos}}, \ and\ \bibinfo {author} {\bibfnamefont
  {M.}~\bibnamefont {R\"ubhausen}},\ }\href {\doibase
  10.1103/PhysRevLett.92.097203} {\bibfield  {journal} {\bibinfo  {journal}
  {Phys. Rev. Lett.}\ }\textbf {\bibinfo {volume} {92}},\ \bibinfo {pages}
  {097203} (\bibinfo {year} {2004})}\BibitemShut {NoStop}%
\bibitem [{\citenamefont {Kim}\ \emph {et~al.}(2002)\citenamefont {Kim},
  \citenamefont {Jung}, \citenamefont {Kim}, \citenamefont {Lee}, \citenamefont
  {Yu}, \citenamefont {Noh},\ and\ \citenamefont {Moritomo}}]{Kim02}%
  \BibitemOpen
  \bibfield  {author} {\bibinfo {author} {\bibfnamefont {M.~W.}\ \bibnamefont
  {Kim}}, \bibinfo {author} {\bibfnamefont {J.~H.}\ \bibnamefont {Jung}},
  \bibinfo {author} {\bibfnamefont {K.~H.}\ \bibnamefont {Kim}}, \bibinfo
  {author} {\bibfnamefont {H.~J.}\ \bibnamefont {Lee}}, \bibinfo {author}
  {\bibfnamefont {J.}~\bibnamefont {Yu}}, \bibinfo {author} {\bibfnamefont
  {T.~W.}\ \bibnamefont {Noh}}, \ and\ \bibinfo {author} {\bibfnamefont
  {Y.}~\bibnamefont {Moritomo}},\ }\href {\doibase
  10.1103/PhysRevLett.89.016403} {\bibfield  {journal} {\bibinfo  {journal}
  {Phys. Rev. Lett.}\ }\textbf {\bibinfo {volume} {89}},\ \bibinfo {pages}
  {016403} (\bibinfo {year} {2002})}\BibitemShut {NoStop}%
\bibitem [{\citenamefont {Kovaleva}\ \emph {et~al.}(2010)\citenamefont
  {Kovaleva}, \citenamefont {Ole\ifmmode~\acute{s}\else \'{s}\fi{}},
  \citenamefont {Balbashov}, \citenamefont {Maljuk}, \citenamefont {Argyriou},
  \citenamefont {Khaliullin},\ and\ \citenamefont {Keimer}}]{Kov10}%
  \BibitemOpen
  \bibfield  {author} {\bibinfo {author} {\bibfnamefont {N.~N.}\ \bibnamefont
  {Kovaleva}}, \bibinfo {author} {\bibfnamefont {A.~M.}\ \bibnamefont
  {Ole\ifmmode~\acute{s}\else \'{s}\fi{}}}, \bibinfo {author} {\bibfnamefont
  {A.~M.}\ \bibnamefont {Balbashov}}, \bibinfo {author} {\bibfnamefont
  {A.}~\bibnamefont {Maljuk}}, \bibinfo {author} {\bibfnamefont {D.~N.}\
  \bibnamefont {Argyriou}}, \bibinfo {author} {\bibfnamefont {G.}~\bibnamefont
  {Khaliullin}}, \ and\ \bibinfo {author} {\bibfnamefont {B.}~\bibnamefont
  {Keimer}},\ }\href {\doibase 10.1103/PhysRevB.81.235130} {\bibfield
  {journal} {\bibinfo  {journal} {Phys. Rev. B}\ }\textbf {\bibinfo {volume}
  {81}},\ \bibinfo {pages} {235130} (\bibinfo {year} {2010})}\BibitemShut
  {NoStop}%
\bibitem [{\citenamefont {Snamina}\ and\ \citenamefont
  {Ole{\'{s}}}(2019)}]{Sna19}%
  \BibitemOpen
  \bibfield  {author} {\bibinfo {author} {\bibfnamefont {M.}~\bibnamefont
  {Snamina}}\ and\ \bibinfo {author} {\bibfnamefont {A.~M.}\ \bibnamefont
  {Ole{\'{s}}}},\ }\href {\doibase 10.1088/1367-2630/aaf0d5} {\bibfield
  {journal} {\bibinfo  {journal} {New Journal of Physics}\ }\textbf {\bibinfo
  {volume} {21}},\ \bibinfo {pages} {023018} (\bibinfo {year}
  {2019})}\BibitemShut {NoStop}%
\bibitem [{\citenamefont {Kune\v{s}}\ \emph {et~al.}(2010)\citenamefont
  {Kune\v{s}}, \citenamefont {Leonov}, \citenamefont {Kollar}, \citenamefont
  {Byczuk}, \citenamefont {Anisimov},\ and\ \citenamefont {Vollhardt}}]{Kun10}%
  \BibitemOpen
  \bibfield  {author} {\bibinfo {author} {\bibfnamefont {J.}~\bibnamefont
  {Kune\v{s}}}, \bibinfo {author} {\bibfnamefont {I.}~\bibnamefont {Leonov}},
  \bibinfo {author} {\bibfnamefont {M.}~\bibnamefont {Kollar}}, \bibinfo
  {author} {\bibfnamefont {K.}~\bibnamefont {Byczuk}}, \bibinfo {author}
  {\bibfnamefont {V.}~\bibnamefont {Anisimov}}, \ and\ \bibinfo {author}
  {\bibfnamefont {D.}~\bibnamefont {Vollhardt}},\ }\href {\doibase
  10.1140/epjst/e2010-01209-0} {\bibfield  {journal} {\bibinfo  {journal} {Eur.
  Phys. J. Special Topics}\ }\textbf {\bibinfo {volume} {180}},\ \bibinfo
  {pages} {5} (\bibinfo {year} {2010})}\BibitemShut {NoStop}%
\bibitem [{\citenamefont {Feiner}\ and\ \citenamefont {Ole\'s}(2005)}]{Fei05}%
  \BibitemOpen
  \bibfield  {author} {\bibinfo {author} {\bibfnamefont {L.~F.}\ \bibnamefont
  {Feiner}}\ and\ \bibinfo {author} {\bibfnamefont {A.~M.}\ \bibnamefont
  {Ole\'s}},\ }\href {\doibase 10.1103/PhysRevB.71.144422} {\bibfield
  {journal} {\bibinfo  {journal} {Phys. Rev. B}\ }\textbf {\bibinfo {volume}
  {71}},\ \bibinfo {pages} {144422} (\bibinfo {year} {2005})}\BibitemShut
  {NoStop}%
\bibitem [{\citenamefont {Ole\ifmmode~\acute{s}\else \'{s}\fi{}}\ \emph
  {et~al.}(2000)\citenamefont {Ole\ifmmode~\acute{s}\else \'{s}\fi{}},
  \citenamefont {Feiner},\ and\ \citenamefont {Zaanen}}]{Ole00}%
  \BibitemOpen
  \bibfield  {author} {\bibinfo {author} {\bibfnamefont {A.~M.}\ \bibnamefont
  {Ole\ifmmode~\acute{s}\else \'{s}\fi{}}}, \bibinfo {author} {\bibfnamefont
  {L.~F.}\ \bibnamefont {Feiner}}, \ and\ \bibinfo {author} {\bibfnamefont
  {J.}~\bibnamefont {Zaanen}},\ }\href {\doibase 10.1103/PhysRevB.61.6257}
  {\bibfield  {journal} {\bibinfo  {journal} {Phys. Rev. B}\ }\textbf {\bibinfo
  {volume} {61}},\ \bibinfo {pages} {6257} (\bibinfo {year}
  {2000})}\BibitemShut {NoStop}%
\bibitem [{\citenamefont {Brzezicki}\ \emph {et~al.}(2012)\citenamefont
  {Brzezicki}, \citenamefont {Dziarmaga},\ and\ \citenamefont
  {Ole\ifmmode~\acute{s}\else \'{s}\fi{}}}]{Brz12}%
  \BibitemOpen
  \bibfield  {author} {\bibinfo {author} {\bibfnamefont {W.}~\bibnamefont
  {Brzezicki}}, \bibinfo {author} {\bibfnamefont {J.}~\bibnamefont
  {Dziarmaga}}, \ and\ \bibinfo {author} {\bibfnamefont {A.~M.}\ \bibnamefont
  {Ole\ifmmode~\acute{s}\else \'{s}\fi{}}},\ }\href {\doibase
  10.1103/PhysRevLett.109.237201} {\bibfield  {journal} {\bibinfo  {journal}
  {Phys. Rev. Lett.}\ }\textbf {\bibinfo {volume} {109}},\ \bibinfo {pages}
  {237201} (\bibinfo {year} {2012})}\BibitemShut {NoStop}%
\bibitem [{\citenamefont {Czarnik}\ \emph {et~al.}(2017)\citenamefont
  {Czarnik}, \citenamefont {Dziarmaga},\ and\ \citenamefont
  {Ole\ifmmode~\acute{s}\else \'{s}\fi{}}}]{Cza17}%
  \BibitemOpen
  \bibfield  {author} {\bibinfo {author} {\bibfnamefont {P.}~\bibnamefont
  {Czarnik}}, \bibinfo {author} {\bibfnamefont {J.}~\bibnamefont {Dziarmaga}},
  \ and\ \bibinfo {author} {\bibfnamefont {A.~M.}\ \bibnamefont
  {Ole\ifmmode~\acute{s}\else \'{s}\fi{}}},\ }\href {\doibase
  10.1103/PhysRevB.96.014420} {\bibfield  {journal} {\bibinfo  {journal} {Phys.
  Rev. B}\ }\textbf {\bibinfo {volume} {96}},\ \bibinfo {pages} {014420}
  (\bibinfo {year} {2017})}\BibitemShut {NoStop}%
\bibitem [{\citenamefont {Bieniasz}\ \emph {et~al.}(2018)\citenamefont
  {Bieniasz}, \citenamefont {Wrzosek}, \citenamefont {Ole\'s},\ and\
  \citenamefont {Wohlfeld}}]{Bie19}%
  \BibitemOpen
  \bibfield  {author} {\bibinfo {author} {\bibfnamefont {K.}~\bibnamefont
  {Bieniasz}}, \bibinfo {author} {\bibfnamefont {P.}~\bibnamefont {Wrzosek}},
  \bibinfo {author} {\bibfnamefont {A.~M.}\ \bibnamefont {Ole\'s}}, \ and\
  \bibinfo {author} {\bibfnamefont {K.}~\bibnamefont {Wohlfeld}},\ }\href@noop
  {} {\bibfield  {journal} {\bibinfo  {journal} {arXiv:1809.07120}\ } (\bibinfo
  {year} {2018})}\BibitemShut {NoStop}%
\bibitem [{\citenamefont {Berciu}(2006)}]{Ber06}%
  \BibitemOpen
  \bibfield  {author} {\bibinfo {author} {\bibfnamefont {M.}~\bibnamefont
  {Berciu}},\ }\href {\doibase 10.1103/PhysRevLett.97.036402} {\bibfield
  {journal} {\bibinfo  {journal} {Phys. Rev. Lett.}\ }\textbf {\bibinfo
  {volume} {97}},\ \bibinfo {pages} {036402} (\bibinfo {year}
  {2006})}\BibitemShut {NoStop}%
\bibitem [{\citenamefont {Marchand}\ \emph {et~al.}(2010)\citenamefont
  {Marchand}, \citenamefont {De~Filippis}, \citenamefont {Cataudella},
  \citenamefont {Berciu}, \citenamefont {Nagaosa}, \citenamefont {Prokof'ev},
  \citenamefont {Mishchenko},\ and\ \citenamefont {Stamp}}]{Mar10}%
  \BibitemOpen
  \bibfield  {author} {\bibinfo {author} {\bibfnamefont {D.~J.~J.}\
  \bibnamefont {Marchand}}, \bibinfo {author} {\bibfnamefont {G.}~\bibnamefont
  {De~Filippis}}, \bibinfo {author} {\bibfnamefont {V.}~\bibnamefont
  {Cataudella}}, \bibinfo {author} {\bibfnamefont {M.}~\bibnamefont {Berciu}},
  \bibinfo {author} {\bibfnamefont {N.}~\bibnamefont {Nagaosa}}, \bibinfo
  {author} {\bibfnamefont {N.~V.}\ \bibnamefont {Prokof'ev}}, \bibinfo {author}
  {\bibfnamefont {A.~S.}\ \bibnamefont {Mishchenko}}, \ and\ \bibinfo {author}
  {\bibfnamefont {P.~C.~E.}\ \bibnamefont {Stamp}},\ }\href {\doibase
  10.1103/PhysRevLett.105.266605} {\bibfield  {journal} {\bibinfo  {journal}
  {Phys. Rev. Lett.}\ }\textbf {\bibinfo {volume} {105}},\ \bibinfo {pages}
  {266605} (\bibinfo {year} {2010})}\BibitemShut {NoStop}%
\bibitem [{\citenamefont {Berciu}\ and\ \citenamefont {Fehske}(2011)}]{Ber11}%
  \BibitemOpen
  \bibfield  {author} {\bibinfo {author} {\bibfnamefont {M.}~\bibnamefont
  {Berciu}}\ and\ \bibinfo {author} {\bibfnamefont {H.}~\bibnamefont
  {Fehske}},\ }\href {\doibase 10.1103/PhysRevB.84.165104} {\bibfield
  {journal} {\bibinfo  {journal} {Phys. Rev. B}\ }\textbf {\bibinfo {volume}
  {84}},\ \bibinfo {pages} {165104} (\bibinfo {year} {2011})}\BibitemShut
  {NoStop}%
\bibitem [{\citenamefont {Ebrahimnejad}\ \emph {et~al.}(2016)\citenamefont
  {Ebrahimnejad}, \citenamefont {Sawatzky},\ and\ \citenamefont
  {Berciu}}]{Ebr15}%
  \BibitemOpen
  \bibfield  {author} {\bibinfo {author} {\bibfnamefont {H.}~\bibnamefont
  {Ebrahimnejad}}, \bibinfo {author} {\bibfnamefont {G.~A.}\ \bibnamefont
  {Sawatzky}}, \ and\ \bibinfo {author} {\bibfnamefont {M.}~\bibnamefont
  {Berciu}},\ }\href {\doibase 10.1088/0953-8984/28/10/105603} {\bibfield
  {journal} {\bibinfo  {journal} {J. Phys.: Cond. Mat.}\ }\textbf {\bibinfo
  {volume} {28}},\ \bibinfo {pages} {105603} (\bibinfo {year}
  {2016})}\BibitemShut {NoStop}%
\bibitem [{\citenamefont {Bieniasz}\ \emph {et~al.}(2016)\citenamefont
  {Bieniasz}, \citenamefont {Berciu}, \citenamefont {Daghofer},\ and\
  \citenamefont {Ole\'s}}]{Bie16}%
  \BibitemOpen
  \bibfield  {author} {\bibinfo {author} {\bibfnamefont {K.}~\bibnamefont
  {Bieniasz}}, \bibinfo {author} {\bibfnamefont {M.}~\bibnamefont {Berciu}},
  \bibinfo {author} {\bibfnamefont {M.}~\bibnamefont {Daghofer}}, \ and\
  \bibinfo {author} {\bibfnamefont {A.~M.}\ \bibnamefont {Ole\'s}},\ }\href
  {\doibase 10.1103/PhysRevB.94.085117} {\bibfield  {journal} {\bibinfo
  {journal} {Phys. Rev. B}\ }\textbf {\bibinfo {volume} {94}},\ \bibinfo
  {pages} {085117} (\bibinfo {year} {2016})}\BibitemShut {NoStop}%
\bibitem [{\citenamefont {Bieniasz}\ \emph {et~al.}(2017)\citenamefont
  {Bieniasz}, \citenamefont {Berciu},\ and\ \citenamefont {Ole\'s}}]{Bie17}%
  \BibitemOpen
  \bibfield  {author} {\bibinfo {author} {\bibfnamefont {K.}~\bibnamefont
  {Bieniasz}}, \bibinfo {author} {\bibfnamefont {M.}~\bibnamefont {Berciu}}, \
  and\ \bibinfo {author} {\bibfnamefont {A.~M.}\ \bibnamefont {Ole\'s}},\
  }\href {\doibase 10.1103/PhysRevB.95.235153} {\bibfield  {journal} {\bibinfo
  {journal} {Phys. Rev. B}\ }\textbf {\bibinfo {volume} {95}},\ \bibinfo
  {pages} {235153} (\bibinfo {year} {2017})}\BibitemShut {NoStop}%
\bibitem [{\citenamefont {Brzezicki}\ \emph {et~al.}(2013)\citenamefont
  {Brzezicki}, \citenamefont {Dziarmaga},\ and\ \citenamefont
  {Ole\'s}}]{Brz13}%
  \BibitemOpen
  \bibfield  {author} {\bibinfo {author} {\bibfnamefont {W.}~\bibnamefont
  {Brzezicki}}, \bibinfo {author} {\bibfnamefont {J.}~\bibnamefont
  {Dziarmaga}}, \ and\ \bibinfo {author} {\bibfnamefont {A.~M.}\ \bibnamefont
  {Ole\'s}},\ }\href {\doibase 10.1103/PhysRevB.87.064407} {\bibfield
  {journal} {\bibinfo  {journal} {Phys. Rev. B}\ }\textbf {\bibinfo {volume}
  {87}},\ \bibinfo {pages} {064407} (\bibinfo {year} {2013})}\BibitemShut
  {NoStop}%
\bibitem [{\citenamefont {Liechtenstein}\ \emph {et~al.}(1995)\citenamefont
  {Liechtenstein}, \citenamefont {Anisimov},\ and\ \citenamefont
  {Zaanen}}]{Lie95}%
  \BibitemOpen
  \bibfield  {author} {\bibinfo {author} {\bibfnamefont {A.~I.}\ \bibnamefont
  {Liechtenstein}}, \bibinfo {author} {\bibfnamefont {V.~I.}\ \bibnamefont
  {Anisimov}}, \ and\ \bibinfo {author} {\bibfnamefont {J.}~\bibnamefont
  {Zaanen}},\ }\href {\doibase 10.1103/PhysRevB.52.R5467} {\bibfield  {journal}
  {\bibinfo  {journal} {Phys. Rev. B}\ }\textbf {\bibinfo {volume} {52}},\
  \bibinfo {pages} {R5467} (\bibinfo {year} {1995})}\BibitemShut {NoStop}%
\bibitem [{\citenamefont {Kataoka}(2004)}]{Kat04}%
  \BibitemOpen
  \bibfield  {author} {\bibinfo {author} {\bibfnamefont {M.}~\bibnamefont
  {Kataoka}},\ }\href {\doibase 10.1143/JPSJ.73.1326} {\bibfield  {journal}
  {\bibinfo  {journal} {J. Phys. Soc. Jpn.}\ }\textbf {\bibinfo {volume}
  {73}},\ \bibinfo {pages} {1326} (\bibinfo {year} {2004})}\BibitemShut
  {NoStop}%
\bibitem [{\citenamefont {Pavarini}\ \emph {et~al.}(2008)\citenamefont
  {Pavarini}, \citenamefont {Koch},\ and\ \citenamefont
  {Lichtenstein}}]{Pav08}%
  \BibitemOpen
  \bibfield  {author} {\bibinfo {author} {\bibfnamefont {E.}~\bibnamefont
  {Pavarini}}, \bibinfo {author} {\bibfnamefont {E.}~\bibnamefont {Koch}}, \
  and\ \bibinfo {author} {\bibfnamefont {A.~I.}\ \bibnamefont {Lichtenstein}},\
  }\href {\doibase 10.1103/PhysRevLett.101.266405} {\bibfield  {journal}
  {\bibinfo  {journal} {Phys. Rev. Lett.}\ }\textbf {\bibinfo {volume} {101}},\
  \bibinfo {pages} {266405} (\bibinfo {year} {2008})}\BibitemShut {NoStop}%
\bibitem [{\citenamefont {Leonov}\ \emph {et~al.}(2010)\citenamefont {Leonov},
  \citenamefont {Korotin}, \citenamefont {Binggeli}, \citenamefont {Anisimov},\
  and\ \citenamefont {Vollhardt}}]{Leo10}%
  \BibitemOpen
  \bibfield  {author} {\bibinfo {author} {\bibfnamefont {I.}~\bibnamefont
  {Leonov}}, \bibinfo {author} {\bibfnamefont {D.}~\bibnamefont {Korotin}},
  \bibinfo {author} {\bibfnamefont {N.}~\bibnamefont {Binggeli}}, \bibinfo
  {author} {\bibfnamefont {V.~I.}\ \bibnamefont {Anisimov}}, \ and\ \bibinfo
  {author} {\bibfnamefont {D.}~\bibnamefont {Vollhardt}},\ }\href {\doibase
  10.1103/PhysRevB.81.075109} {\bibfield  {journal} {\bibinfo  {journal} {Phys.
  Rev. B}\ }\textbf {\bibinfo {volume} {81}},\ \bibinfo {pages} {075109}
  (\bibinfo {year} {2010})}\BibitemShut {NoStop}%
\bibitem [{\citenamefont {Binggeli}\ and\ \citenamefont
  {Altarelli}(2004)}]{Bin04}%
  \BibitemOpen
  \bibfield  {author} {\bibinfo {author} {\bibfnamefont {N.}~\bibnamefont
  {Binggeli}}\ and\ \bibinfo {author} {\bibfnamefont {M.}~\bibnamefont
  {Altarelli}},\ }\href {\doibase 10.1103/PhysRevB.70.085117} {\bibfield
  {journal} {\bibinfo  {journal} {Phys. Rev. B}\ }\textbf {\bibinfo {volume}
  {70}},\ \bibinfo {pages} {085117} (\bibinfo {year} {2004})}\BibitemShut
  {NoStop}%
\bibitem [{\citenamefont {Pavarini}\ and\ \citenamefont {Koch}(2010)}]{Pav10}%
  \BibitemOpen
  \bibfield  {author} {\bibinfo {author} {\bibfnamefont {E.}~\bibnamefont
  {Pavarini}}\ and\ \bibinfo {author} {\bibfnamefont {E.}~\bibnamefont
  {Koch}},\ }\href {\doibase 10.1103/PhysRevLett.104.086402} {\bibfield
  {journal} {\bibinfo  {journal} {Phys. Rev. Lett.}\ }\textbf {\bibinfo
  {volume} {104}},\ \bibinfo {pages} {086402} (\bibinfo {year}
  {2010})}\BibitemShut {NoStop}%
\end{thebibliography}
%merlin.mbs apsrev4-1.bst 2010-07-25 4.21a (PWD, AO, DPC) hacked
%Control: key (0)
%Control: author (72) initials jnrlst
%Control: editor formatted (1) identically to author
%Control: production of article title (-1) disabled
%Control: page (0) single
%Control: year (1) truncated
%Control: production of eprint (0) enabled
%

\end{document}